\documentclass[twocolumn,natbib,tighten,twocolappendix,times]{aastex63}

\usepackage{xspace}
\usepackage{float}
\usepackage{xcolor,color}
\usepackage{multirow}
\usepackage{amsmath}
\usepackage{enumitem}
\usepackage{longtable}

\defcitealias{Robotham2015}{RO15}
\defcitealias{Akritas1996}{AB96}

\newcommand{\Sec}[1]{{\protect\hyperref[sec:#1]{Section~\ref*{sec:#1}}}}
\newcommand{\Secs}[2]{{\protect\hyperref[sec:#1]{Sections~\ref*{sec:#1}}~and~\ref{sec:#2}}}
\newcommand{\Fig}[1]{{\protect\hyperref[fig:#1]{Figure~\ref*{fig:#1}}}}
\newcommand{\Equ}[1]{{\protect\hyperref[equ:#1]{Equation~\ref*{equ:#1}}}}

\newcommand{\Tab}[1]{{\protect\hyperref[tab:#1]{Table~\ref*{tab:#1}}}}
\newcommand{\Tabs}[2]{{\protect\hyperref[tab:#1]{Tables~\ref*{tab:#1}}}~and~\ref{tab:#2}}
\newcommand{\App}[1]{{\protect\hyperref[app:#1]{Appendix~\ref*{app:#1}}}}

\newcommand{\HI}{H{\footnotesize I}\xspace}

\newcommand{\Ha}{H\ensuremath{\alpha}\xspace}

\newcommand{\wunits}[2]{\ensuremath{#1\,\text{#2}}}

\received{November 25, 2020}
\revised{January 25, 2021}
\accepted{March 2, 2021}
\submitjournal{The Astrophysical Journal}

\shorttitle{Intrinsic Scatter}
\shortauthors{Stone, Courteau, and Arora}

\begin{document}

\title{The Intrinsic Scatter of Galaxy Scaling Relations}

\correspondingauthor{Connor Stone}
\email{connor.stone@queensu.ca}

\author[0000-0002-9086-6398]{Connor Stone}
\affiliation{Department of Physics, Engineering Physics and Astronomy,
  Queen{'}s University,
  Kingston, ON K7L 3N6, Canada}

\author[0000-0002-8597-6277]{St{\'e}phane Courteau}
\affiliation{Department of Physics, Engineering Physics and Astronomy,
  Queen{'}s University,
  Kingston, ON K7L 3N6, Canada}
  
  \author[0000-0002-3929-9316]{Nikhil Arora}
\affiliation{Department of Physics, Engineering Physics and Astronomy,
  Queen{'}s University,
  Kingston, ON K7L 3N6, Canada}

\begin{abstract}
  We present a compendium of disk galaxy scaling relations and a detailed characterization of their intrinsic scatter.
  Observed scaling relations are typically characterized by their slope, intercept, and scatter; however, these parameters are a mixture of observational errors and astrophysical processes. 
  We introduce a novel Bayesian framework for computing the intrinsic scatter of scaling relations that accounts for nonlinear error propagation and covariant uncertainties.  
  Bayesian intrinsic scatters are $\sim25$ percent more accurate than those obtained with a first-order classical method, which systematically underestimates the true intrinsic scatter.
  Structural galaxy scaling relations based on velocity ($V_{23.5}$), size ($R_{23.5}$), luminosity ($L_{23.5}$), colour ($g-z$), central stellar surface density ($\Sigma_1$), stellar mass ($M_{*}$), dynamical mass ($M_{\rm dyn}$), stellar angular momentum ($j_{*}$), and dynamical angular momentum ($j_{\rm dyn}$), are examined to demonstrate the power and importance of the Bayesian formalism.
  Our analysis is based on a diverse selection of over 1000 late-type galaxies from the Photometry and Rotation Curve Observations from Extragalactic Surveys compilation with deep optical photometry and extended rotation curves. 
  We determine the tightest relation for each parameter by intrinsic orthogonal scatter, finding $M_{*}-V_{23.5}$, $R_{23.5}-j_{*}$, and $L_{23.5}-j_{\rm dyn}$ to be especially tight.
  The scatter of the $R_{23.5}-L_{23.5}$, $V_{23.5} - (g-z)$, and $R_{23.5}-j_{\rm dyn}$ relations is mostly intrinsic, making them ideal for galaxy formation and evolutionary studies.
  Our code to compute the Bayesian intrinsic scatter of any scaling relation is also presented.
  We quantify the correlated nature of many uncertainties in galaxy scaling relations and scrutinize the uncertain nature of disk inclination corrections and their effect on scatter estimates.
\end{abstract}


\section{Introduction} \label{sec:intro}

Scaling relations are a critical component of any characterization and understanding of galaxy formation and evolutionary scenarios.
For instance, a necessary benchmark for the success and fine-tuning of hydrodynamics or semi-analytic models of galaxies is the degree to which they reproduce the slope and scatter of known scaling relations for observed galaxies~\citep{Steinmetz1999,Brook2012b,Scannapieco2012,Knebe2018,Lagos2018}.
In addition, one can describe the evolution of galaxies with time (redshift) in terms of scaling relations~\citep{Mo1998,Hopkins2009,Peng2010b,vanderWel2014,Mowla2019}.
Scaling relations of observed galaxies and their residuals are also valuable for estimating galaxy distances~\citep{Tully1977,Jacoby1992,Willick1999,Sakai2000,Kourkchi2020}
and fine-tuning structural parameter corrections~\citep{Tully1985,Giovanelli1994,Willick1997,Courteau2007,Giovanelli2013}.
They can also drive the discovery of new physical relationships~\citep{Bender1993,Woo2008,Beifiori2012,Ellison2020}.

Galaxy scaling relations are typically linear in log scale and are therefore characterized by three quantities: slope, intercept, and scatter.
Cold Dark Matter ($\Lambda$CDM) galaxy formation models that match both the abundance and size distribution of galaxies can reproduce some galaxy scaling relations such as velocity-luminosity, also known as the Tully-Fisher relation \citep[TFR;][]{Tully1977}, with reasonable accuracy \citep{Dutton2011b,Brook2012b,Ferrero2017}.
Simultaneously matching multiple scaling relations is, however, a challenging task~\citep{Dutton2011b,Trujillo-Gomez2011}.
Small changes to a numerical model can indeed result in significant changes to slope, intercept, and/or scatter for one or more scaling relations simultaneously \citep{Dutton2010,Brook2012, Kim2014,Schaller2015}.

Scaling relation slopes and scatters reported in the literature can vary greatly as a result of the many different selection functions, systematic errors from heterogeneous reduction and analysis methods, observational errors, bandpass sensitivities, and other effects (see \Sec{discussion}). 
Short of being able to correct for all biases and systematic differences, attempts to compare empirical scaling relations with theoretical models of galaxy structure require at least that most (tractable) observational errors be removed \citep{Strauss1995,Pizagno2005,Pizagno2007,Saintonge2011,Lelli2017}.
In this work, we present a detailed framework for the simultaneous derivation of the intrinsic scatter estimates for numerous galaxy scaling relations.

The intrinsic slope, intercept and total scatter of a scaling relation can be found to arbitrary precision (limit of random error goes to zero) by fitting a model to a large unbiased galaxy sample. 
However, the intrinsic scatter must also be inferred by modeling uncertainties.
Due to the complex and heteroscedastic nature of observational uncertainties in astronomy, accounting for their impact in empirical studies can be challenging~\citep{Takashi1990,Andreon2013}.
Two broad paths are typically explored to connect galaxy observations and models.
One path consists of generating mock observations of a model by introducing uncertainties tuned to a specific observational campaign~\citep{Jonsson2010,Snyder2015,Torrey2015,Bottrell2017,Yung2019}.
This technique has the advantage of being flexible; once the mock observations are complete, one can apply standard analysis pipelines and produce any desired quantity for a large sample size.
The results are however nontransferable between different models and observations, and so must be redone for each model-observation pair.
The second path accounts directly for uncertainties and effectively subtracts their effects from an analysis of observational data in order to arrive at intrinsic quantities.
This second method is in principle preferable since physically meaningful results are directly extracted and can instantly be compared with any model.
The main drawback of this approach is the complexity in accounting for sample completeness, bias, and uncertainty in the data.

The TFR is a case where the intrinsic scatter has been inferred from observations~\citep{Bernstein1994,Gnedin2007,Pizagno2007,Reyes2011,Saintonge2011}, and likewise for the radial acceleration relation~\citep{McGaugh2004,McGaugh2016,Lelli2017}
where intrinsic scatters can differentiate physical models~\citep{McGaugh2016,Rong2018,Stone2019}.
However, intrinsic scatter estimates in some of these examples are determined to first order by subtracting the average uncertainty in quadrature.
This ``classical'' analysis (detailed in \Sec{classicalintrinsicscatter}) has notable drawbacks, including the absence of correlated uncertainties and an assumption of linearity in all transformations, not to mention the possibility of returning nonphysical negative scatters.
In contrast, some weak lensing scaling relations are now benefiting from Bayesian estimates of intrinsic scatter that more robustly account for sources of error~\citep{Sereno2015}.

In this work, we present a Bayesian framework for analyzing intrinsic scatter and apply it to a suite of well-known structural and dynamical galaxy scaling relations.
This framework is easily implemented numerically and provides robust measures of intrinsic scatter, even when confronted with certain biases that the classical analysis cannot take into consideration~(\App{testtruncation}).

This paper is organized as follows. 
We begin by modeling the scatter of a general scaling relation in \Sec{intrinsicscatter} and explicitly define the classical and Bayesian procedures for evaluating intrinsic scatters.
\Sec{datasets} then describes the Photometry and Rotation Curve Observations from Extragalactic Surveys (PROBES) compilation that is used in our analysis; it represents the largest collection of deep long slit spectroscopic rotation curves coupled with deep multiband photometry to date.
\Sec{results} presents the fit results and intrinsic scatter measurements following the outlined analysis techniques.
\Sec{discussion} performs a detailed literature comparison with our results and many others, including a comparison of intrinsic scatters where available.
We expound in \Sec{conclusions} on robust measures of intrinsic scatter as a powerful metric for validating and improving galaxy evolutionary models.
For more information on the numerical technique, see \App{testtruncation} where we present the numerical method to compute Bayesian intrinsic scatters and tests of the technique's stability.
\App{bayesianintrinsicscattercode} provides a minimalist python implementation of the Bayesian intrinsic scatter code for ease of use by other authors.
In \Sec{inclinationcorrections} we explore the uncertain nature of inclination corrections and the importance of a multi-scaling relation analysis to determine their true form.

\section{Intrinsic Scatter Model}
\label{sec:intrinsicscatter}

We have noted that any (linear) scaling relation can be represented by a slope, intercept, and scatter.  
This section pertains specifically to the characterization of a scaling relation's intrinsic scatter.
Modeling the scatter of a scaling relation involves clearly defining the source of each perturbation on a data point and tracking their effect on the position of this point in a given mathematical transformation.
We express the mathematical model used to define intrinsic scatter and its relation to observational uncertainties.
In a classical framework, the intrinsic scatter is calculated by averaging the observational uncertainties to first order and subtracting them in quadrature from the total observed scatter.
In a Bayesian framework, a posterior is constructed for the intrinsic scatter by marginalizing over all observational uncertainties.
Both frameworks are reviewed below.

\subsection{Scatter Model for Scaling Relations}
\label{sec:scattermodel}

We first assume a vector of variables $\theta$ that describes the measurements taken of an object (e.g. a galaxy).
A scaling relation is constructed using the functions $X(\theta)$ and $Y(\theta)$ for each axis.
Some values in $\theta$ may not be used in the $X$ or $Y$ functions and some may be used by both (likely leading to covariance); $\theta$ simply represents all necessary inputs to compute the relation.
The scaling relation is a function $f$ such that $f(X(\theta)) = Y(\theta)$ for a ``perfect'' relation.
Nature does not produce perfect (scatter-free) relations though; instead, every object (galaxy) will suffer a  perturbation from the relation drawn from an intrinsic scatter distribution.
Since the exact nature of the intrinsic scatter cannot be recovered exactly, its form is assumed to be a normal distribution, which scatters about the $y$-axis.
There is some loss of generality in the assumption of normal distribution; however there is no compelling reason to suspect otherwise.
Thus our scaling relation can be written $f(X(\theta)) = Y(\theta) + \epsilon$ where $\epsilon$ is the perturbation drawn from $\mathcal{N}(\epsilon | 0,\sigma_i^2)$ where $\mathcal{N}$ is a normal distribution probability density function (pdf) and $\sigma_i$ is the intrinsic scatter.

The quantities $\theta$ cannot be measured perfectly, and we measure instead $\phi = \theta + \tilde{\theta}$ where $\tilde{\theta}$ is some perturbation drawn from $P(\tilde{\theta}|\sigma_{\phi})$ where $\sigma_{\phi}$ are the uncertainties for each variable.
For each measured quantity, we have access to the measurement ($\phi$) and the uncertainty ($\sigma_{\phi}$); the specific intermediate $\tilde{\theta}$ is unknown though its distribution is known.

The residuals from a scaling relation can then be decomposed into their measurement uncertainty and intrinsic scatter components.
Forward residuals are represented as $R = Y(\phi) - f(X(\phi))$ which after decomposition look like $R = Y(\theta + \tilde{\theta}) - f(X(\theta + \tilde{\theta})) + \epsilon$.
Thus the residual $R$ can be measured, but $\theta,~\tilde{\theta},$ and $\epsilon$ are unknown.
Recovering $\sigma_i$, which generates the $\epsilon$ distribution, can be achieved in various ways as we address below.
\Sec{classicalintrinsicscatter} presents the classical first-order method and \Sec{bayesianintrinsicscatter} presents our more careful Bayesian technique. 

\subsection{Classical Intrinsic Scatter}
\label{sec:classicalintrinsicscatter}

In the classical framework, the observed scatter ($\sigma_o$) and uncertainty scatter ($\sigma_u$) of a scaling relation are computed to first order and subtracted in quadrature to produce the intrinsic scatter.
Assuming a normal distribution, the observed scatter is computed with a standard deviation of the forward residuals; estimating the scatter due to uncertainty is more complex.
For an individual data point, the uncertainty scatter is computed as:

\begin{equation}
  \begin{aligned}
    \sigma_y^2 &= \sum_i\left(\frac{df}{d\theta_i}\sigma_{\theta_i}\right)^2,
    \label{equ:uncertaintyscatter}
  \end{aligned}
\end{equation}

\noindent where $\sigma_y$ is the $y$-axis scatter, $\theta_i$ is a variable used in computing the relation, $\frac{df}{d\theta_i}$ is the derivative of the relation with respect to variable $\theta_i$, and $\sigma_{\theta_i}$ is the uncertainty on variable $\theta_i$.
In the specific context of galaxy scaling relations, data are highly heteroscedastic, and a procedure for averaging the uncertainties is required.
The average of the variances $\sigma_y^2$ (not the standard deviation $\sigma_y$) gives an unbiased estimator of the scatter due to observational uncertainties:

\begin{equation}
  \begin{aligned}
    \sigma_u^2 &= \frac{1}{N}\sum_y\sigma_y^2,
  \end{aligned}
\end{equation}

\noindent  where $N$ is the total number of observations.
The intrinsic scatter is then computed as $\sigma_i^2 = \sigma_o^2 - \sigma_u^2$. Note that while this formula may produce negative intrinsic scatters, these values must be reported despite the unrealistic value for the estimator to be unbiased on average.
The value of interest is $\sigma_i$, not $\sigma_i^2$, and so a square root must be taken.
In the event that $\sigma_i^2$ is negative, we report a negative $\sigma_i$; the estimator for $\sigma_i$ is therefore $\text{sign}(\sigma_i^2)\sqrt{|\sigma_i^2|}$.
A confidence interval for $\sigma_i$ can be determined using bootstrap sampling~\citep{Efron1992}.

\subsection{Bayesian Intrinsic Scatter}
\label{sec:bayesianintrinsicscatter}

The Bayesian framework involves marginalizing over all observational uncertainties, leaving a pdf for possible intrinsic scatter values.
If $\tilde{\theta}$ were known, we could write $\epsilon = Y(\phi - \tilde{\theta}) - f(X(\phi - \tilde{\theta}))$ for each galaxy and construct a pdf for $\sigma_i$ directly.
Since $\tilde{\theta}$ is unknown (only the distribution from which it is drawn), we construct:
\begin{equation}
  \begin{aligned}
    P(\phi|\sigma_i,\tilde{\theta}) = \mathcal{N}(Y(\phi - \tilde{\theta}) - f(X(\phi - \tilde{\theta}))|0,\sigma_i^2),
    \label{equ:bayespdf}
  \end{aligned}
\end{equation}
\noindent which corresponds to the probability of obtaining the residual ($\epsilon$) given a proposed intrinsic scatter ($\sigma_i$) and perturbation ($\tilde{\theta}$) combination.
A notable aspect of the algorithm presented in \Equ{bayespdf} is the generality of the function $f(X(\phi - \tilde{\theta}))$, such that our Bayesian intrinsic scatters may be computed for any scaling relation for which residuals are measured.
The marginalization over the pdf ($P(\tilde{\theta}|\sigma_{\phi})$) of possible $\tilde{\theta}$ values yields a pdf of $\phi$.
Bayes Theorem is then used to convert the result into a pdf for intrinsic scatter:

\begin{equation}
  \begin{aligned}
    P(\phi|\sigma_i,\sigma_{\phi}) &= \int P(\phi|\sigma_i,\tilde{\theta})P(\tilde{\theta}|\sigma_{\phi})d\tilde{\theta} \\
    P(\sigma_i|\phi,\sigma_{\phi}) &= \frac{P(\phi|\sigma_i,\sigma_{\phi})P(\sigma_i)}{P(\phi)},
    \label{equ:bayesscatter}
  \end{aligned}
\end{equation}

\noindent where $P(\sigma_i)$ is the prior for the intrinsic scatter (we use a flat prior from zero to the total scaling relation scatter) and $P(\phi)$ is the normalization.
The product of all posteriors is used to combine the results for many objects (galaxies). 
While simple, this procedure cannot be computed analytically; instead, numerical techniques must be used to perform the integral in \Equ{bayesscatter}.
\Sec{numericalbayes} presents such a procedure for computing the integral.
Tests demonstrating the robustness of the Bayesian method relative to the classical method with a toy model are also presented in \App{testtruncation}.
In our idealized toy model, the estimates of the Bayesian intrinsic scatter have at minimum a \wunits{25}{\%} relative accuracy improvement over classical estimates; however many factors influence the true error for a real scaling relation.
A code to compute a Bayesian intrinsic scatter is also presented in \App{bayesianintrinsicscattercode}. 

\subsection{Numerical Bayesian Intrinsic Scatters}
\label{sec:numericalbayes}

\begin{figure}[ht]
  \centering
  \includegraphics[width=\columnwidth]{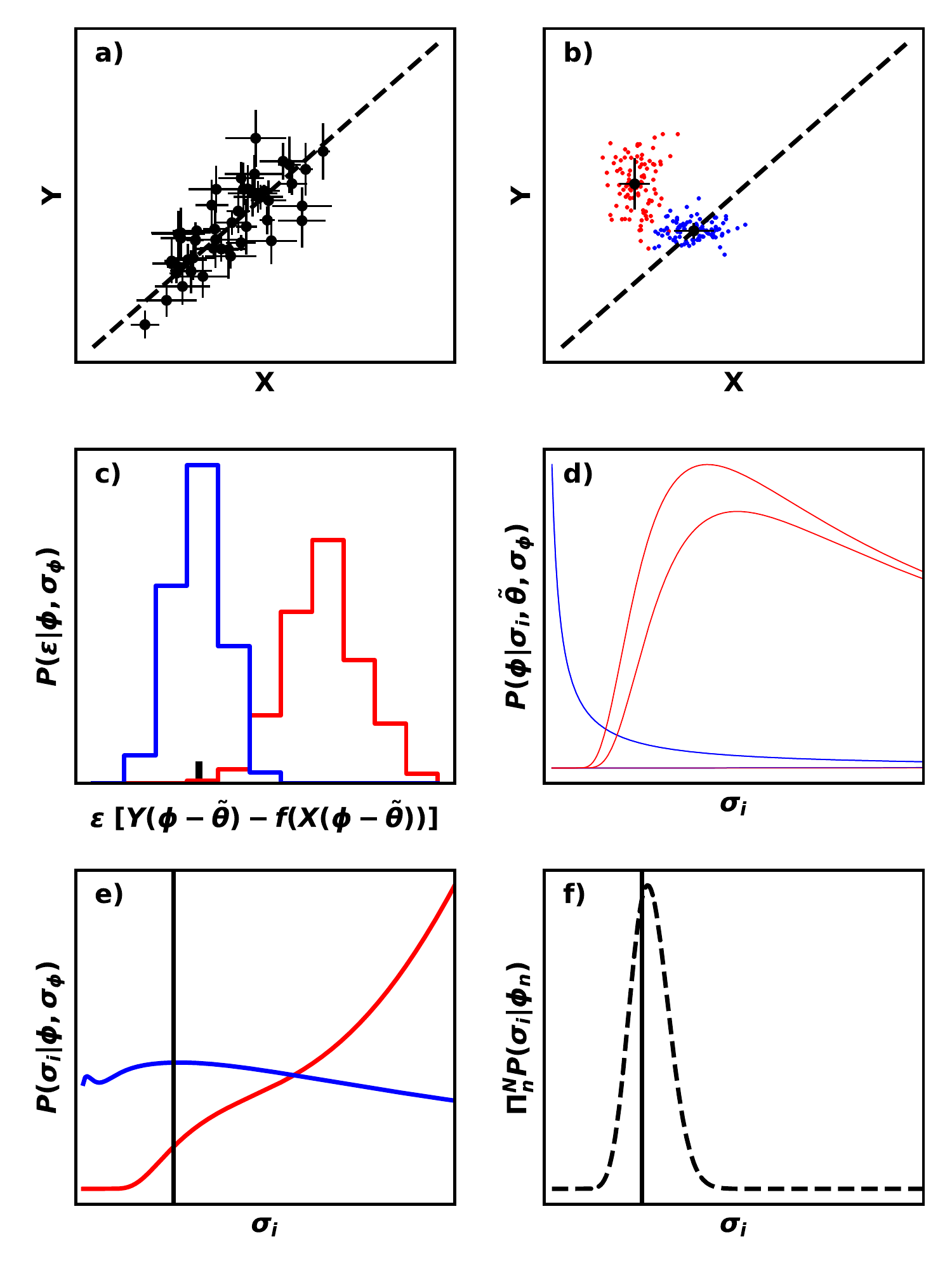}
  \caption{Visualization of the Bayesian intrinsic scatter calculation. a) Mock scaling relation shown with hypothetical heteroscedastic data. b) Sampling $P(\tilde{\theta}|\sigma_{\phi})$ for two example points; red is far from the relation, blue is near the relation. c) Residual pdf of $\epsilon$ for the two sample points, the small black tick on the $x$-axis represents zero. d) Examples of the likelihood for $\phi$. e) Bayesian posterior for $\sigma_i$ for the two sample points; the blue point favours low $\sigma_i$, while the red point favours large $\sigma_i$. f) Product of all Bayesian posteriors. This is the final pdf for $\sigma_i$ given the scaling relation data in part a). The black vertical line in panels e) and f) represents the true intrinsic scatter.}
  \label{fig:Figures/Bayes_Procedure.pdf}
\end{figure}

The procedure for marginalizing over observational uncertainties to get the intrinsic scatter of a scaling relation described in \Sec{bayesianintrinsicscatter} cannot be completed analytically; however, numerical techniques can achieve arbitrary precision.
The integration in \Equ{bayesscatter} can numerically be performed with relative ease, though with a slightly modified procedure.
Depicted in \Fig{Figures/Bayes_Procedure.pdf} are the critical steps of the numerical procedure as performed with a toy model.

The toy model data in \Fig{Figures/Bayes_Procedure.pdf}-a are constructed as follows. 
Five-hundred mock ``galaxies'' are sampled from a two-dimensional Gaussian distribution (only 50 shown for clarity) forming the true distribution of the galaxies ($\theta$) with intrinsic scatter $\sigma_i$ as the forward residual standard deviation.
Each data point is assigned a unique observational uncertainty ($\sigma_{\phi}$) in the $x$-axis and $y$-axis separately by drawing an uncertainty from a uniform distribution over the range $[0.03,0.3]$, thus making the data heteroscedastic.
The data points are then perturbed by sampling normal distributions with variance determined by their assigned observational uncertainty; this simulates the $\tilde{\theta}$ perturbations giving $\phi = \theta + \tilde{\theta}$.
The Bayesian algorithm knows only the final position of the data points ($\phi$) and the assigned observational uncertainty for each data point ($\sigma_{\phi}$).
The data are fit with a BCES bisector algorithm (see \Sec{scalingrelationfits}) and all residuals are computed relative to this.

In \Fig{Figures/Bayes_Procedure.pdf}-b the observational uncertainties ($P(\tilde{\theta}|\sigma_{\phi})$) are numerically integrated over by sampling their distribution 500 times (only 100 shown for clarity).
In \Fig{Figures/Bayes_Procedure.pdf}-c the residuals of the samples form a pdf for the perturbation $\epsilon$ given the observed data point and observational uncertainties.
\Fig{Figures/Bayes_Procedure.pdf}-d shows the likelihood for $\phi$ as a function of possible $\sigma_i$ values.
This is essentially an inverted gaussian distribution, the probability density is plotted by varying $\sigma$ for a fixed x (not the other way around) where x is a single residual value after sampling in \Fig{Figures/Bayes_Procedure.pdf}-b.
In \Fig{Figures/Bayes_Procedure.pdf}-e, the ``integral'' is performed by summing each $\phi$ likelihood at a given $\sigma_i$ value, and Bayes theorem is used to turn this into a pdf for $\sigma_i$.
Only two examples are shown for clarity, one that favours low $\sigma_i$ values and one that favours large $\sigma_i$ values.
Finally, in \Fig{Figures/Bayes_Procedure.pdf}-f, the product of all posteriors is taken to give the posterior for the full dataset in \Fig{Figures/Bayes_Procedure.pdf}-a.

\section{PROBES Sample}
\label{sec:datasets}

We have described a new and powerful method for computing intrinsic scatters for arbitrary scaling relations.
To demonstrate the general usefulness of this algorithm, we will apply it to a compendium of galaxy scaling relations.
We begin by describing the data used in our analysis.

The PROBES compilation combines several rotation curve surveys and homogeneous photometry from the Dark Energy Spectroscopic Instrument Legacy Imaging Survey \citep[DESI-LIS;][]{Dey2019}.
Here we describe the parameter extraction techniques used in this analysis as well as the choices for corrections, uncertainty propagation, and data quality cuts.

\subsection{Light Profile Extraction Using AutoProf}
\label{sec:desidata}

Photometry for our analysis comes from the DESI-LIS, which provides images in the $g-$, $r-$, and $z-$bands for a large ($\approx 14,000$ deg$^2$) section of the sky \citep{Dey2019}.
The intersection of DESI-LIS and PROBES  (\Sec{probesrotationcurves}) totals 1396 galaxies to be used for our analysis, with some removed due to selection cuts described in \Sec{dataqualitycuts}.
Images are processed using our surface photometry package, AutoProf, which is briefly described below.

AutoProf is our Python-based galaxy image isophotal solution pipeline, with functions for center finding, background subtraction, star masking, isophote fitting, and surface brightness profile extraction.
Center finding uses centering methods from the photutils Python package~\citep{Bradley2020} or with a user-defined center (in pixel coordinates).
Background subtraction and star masking are also completed using standard tools from the photutils package.
Star masking is generally turned off by default, but wrappers are included for IRAF star finding~\citep{Tody1986} and the DAO star finder~\citep{Stetson1987}.
The isophote optimization algorithm minimizes the amplitude of low-order fast Fourier transform coefficients~\citep{cooley1965} for flux values evaluated around an isophote, plus a regularization term~\citep{Shalev2014}.
The regularization term penalizes large differences in ellipticity and position angle between adjacent isophotes using the $l_{2}$ norm (the sum of squared differences; also see \citealt{Shalev2014}).
This effectively smooths out the isophotal solution while not setting any explicit boundaries on the difference between adjacent isophotes. 
In some cases, typically for non-axisymmetric features such as bars or strong spiral arms, it is desirable to allow for large variations in ellipticity and/or position angle. 
Multiple isophotes are fit simultaneously; these are selected with geometrically growing radii out to a signal-to-noise ratio (S/N) of $\sim10$ (typically \wunits{23.5}{$z$-mag\,arcsec$^{-2}$}), beyond which the ellipticity and position angle are taken as constant.
To sample a surface brightness profile, the isophotal solution ellipticity and position angle are linearly interpolated allowing for any desired sampling of the image.
Surface brightnesses are taken as the median flux from many sample points around an isophote, and so most foreground stars need not be masked from an image as they will be ignored by the median.
A curve of growth is computed as the integral of the surface brightness profile; the summed flux of all pixels within each isophote is reported as well.
The fluxes are converted to the AB magnitude system\footnote{See \citet{Dey2019} and \url{https://www.legacysurvey.org/dr8/description/} for specifics of the DESI-LIS photometry}.

Isophotal solutions are visually inspected to identify possible failures such as those involving non-axisymmetric features. 
Most PROBES galaxies already have archival photometry available, though for a limited range of photometric bands. 
These (heterogeneous) light profiles were used to validate the accuracy of the automated DESI-LIS photometry.
Galaxies deemed to have failed the visual isophote inspection or deviate in some pathological way from the older PROBES photometry were discarded.
The surface brightness profiles are evaluated out to a typical photometric depth in the $z$ band of roughly \wunits{26}{mag arcsec$^{-2}$} before reaching a cutoff uncertainty of \wunits{0.2}{mag arcsec$^{-2}$}.
AutoProf will be released in a future publication.

\subsection{Input Parameters}
\label{sec:uncertaintymodel}

PROBES is composed of a large set of galactic observations.
These observations take the form of rotation curves, photometry, distances, and morphological types.
Below we describe the nature of the PROBES observations and uncertainty modeling.
These uncertainties will be propagated through all extracted parameters (\Sec{extractedparameters}), necessary for the computation of intrinsic scatters.

\subsubsection{Rotation Curves}\label{sec:probesrotationcurves}

PROBES draws rotation curves from seven different surveys, each with different selection criteria. See \citet{Stone2019} for a brief description of the PROBES compilation or consult the original survey papers for more detail \citep{Mathewson1992,Mathewson1996,Courteau1997,Courteau2000,Lelli2016,Ouellette2017}.

The survey by \citet{Dale1999} is not described in \citet{Stone2019} and was added subsequently to PROBES. This sky spanning sample, also referred to as ``SCII'', provides 522 \Ha rotation curves for galaxies from 52 Abell clusters up to redshifts of $\sim$\wunits{25,000}{km\,s$^{-1}$}. Distances to these galaxies come from a combination of Hubble-Lema\^{i}tre flow and cluster distances. Galaxies are selected from the Abell Rich Cluster Catalog \citep{Abell1989}, favouring those with redshift information available at the time.

Observed velocities are measured through long slit \Ha spectroscopy for all galaxies except those assembled in the \citet{Lelli2016} ``SPARC'' compilation, which has \HI or hybrid \HI / \Ha profiles.
A rotation curve must include a minimum of 10 independent radial points in order to be included in our analysis.
The global recessional velocity is subtracted from each rotation curve by fitting the \citet{Courteau1997} multiparameter model using a regularized, error weighted, least-squares fit~\citep{Tibshirani1996,Zou2005,Friedman2010}.
Some surveys already have subtracted recessional velocities; however, we readjust the velocity centers  with our multiparameter model for consistency (with one exception).
\citet{Lelli2016} presented folded rotation curves (all measurements at positive radii), so no further recessional velocity subtraction is performed.
For those that do require a global velocity subtraction, we compare the \citet{Courteau1997} multiparameter model fits with a simple arctan model (also used in \citealt{Courteau1997}) and find a good agreement with \wunits{3}{km\,s$^{-1}$} scatter.
This average scatter is included in our uncertainty model as a random global shift in the rotation curve for the Bayesian analysis. 
For the classical analysis, it is added in quadrature to the velocity measurement uncertainties. 
Most surveys report both velocity measurements and uncertainties for each data point; if uncertainties are not provided, we use the standard deviation of the residuals from the \citet{Courteau1997} multiparameter model fit as the uncertainty (this is an upper bound for the average uncertainty).

\subsubsection{Photometry}
\label{sec:uncertaintymodelphotometry}

The photometry extraction procedure described in \Sec{desidata} also generates an uncertainty for all measurements.
For surface brightness values, the uncertainty is determined by taking the half $16-84$ interquartile range and dividing by the square root of the number of samples.
The half $16-84$ interquartile range is used instead of the standard deviation as it is more robust to outliers (pixels affected by foreground stars), but limits to the same value for normally distributed data.
These uncertainties are calculated in linear flux space and converted to mag\,arcsec$^{-2}$ with $\left|2.5\frac{\sigma_f}{f\cdot\ln(10)}\right|$ where $f$ is the flux and $\sigma_f$ is the uncertainty.
We also include a global photometric uncertainty of \wunits{0.02}{mag\,arcsec$^{-2}$} for all surface brightness values to account for uncertainty in background subtraction, flux calibration, and global model differences such as center selection. 
The curve of growth is computed by integrating the surface brightness profile.
The uncertainty for the curve of growth at each point is determined by Monte Carlo sampling many surface brightness profiles (using the uncertainty for each point) and re-integrating.
We then use the half $16-84$ interquartile range of the many samples to determine an uncertainty for each point in the curve of growth.

An ellipticity and position angle profile for each galaxy are also extracted using AutoProf. 
The ellipticity values are used to compute a representative inclination for the outer disk of each galaxy according to \Equ{inclination}. 
Comparison of our ellipticity values with those reported in the original PROBES surveys yields an average ellipticity error, $\sigma_e$, of 0.05, which is used as the uncertainty for all ellipticity values. 
Given non-axisymmetric features in galaxies, we do not estimate ellipticity errors uniquely for each galaxy.
The outer disk ellipticity at approximately \wunits{23.5}{$z$-mag arcsec$^{-2}$} from each profile is used to compute the global inclination of a disk galaxy via:

\begin{equation}\label{equ:inclination}
  \begin{aligned}
    \cos^2(i) = \frac{q^2 - q_0^2}{1-q_0^2}
  \end{aligned}
\end{equation}

\noindent where $q$ is the axis ratio ($b/a$); $q_0$ is the galaxy intrinsic thickness ($c/a$); $a$, $b$, $c$, are the principal (semi-)axes of the galaxy; and $i$ is the galaxy inclination~\citep{Hubble1926}.
Since $q_0$ cannot be directly measured for each galaxy, a fixed value is assumed for our sample of $q_0 = 0.13$~\citep{Hall2012}.
There is an uncertainty associated with $q_0$ for which many values have been proposed~\citep{Haynes1984,Lambas1992,Mosenkov2015}; we associate an error of $0.05$ with $q_0$.
This somewhat arbitrary uncertainty value spans the range of proposed $q_0$ values. 
The sampling range is restricted to $0.1 < q_0 < 0.23$ so as not to reach unphysical values.  It also corresponds to the range proposed in \citet{Haynes1984}.
Inclination plays a critical role in the analysis of galaxy structure as it is used to correct most of the extracted structural parameters described in \Sec{extractedparameters}.
As such, the choice of $q_0$ can influence the resulting fitted slope and scatter by over \wunits{10}{\%}, though more typically of the order of \wunits{2}{\%}.
We consider our $q_0 = 0.13\pm0.05$ value close enough to the true thickness for our purposes, though further study is severely needed.

Our photometry is also corrected for galactic extinction using the \citet{Schlegel1998} dust map for each of the $grz$ bands extracted using NED\footnote{The NASA/IPAC Extragalactic Database (NED) is operated by the Jet Propulsion Laboratory, California Institute of Technology, under contract with the National Aeronautics and Space Administration.}.
An uncertainty of \wunits{0.02}{mag} is assumed for all extinction values.
We do not apply any $K$-correction as all PROBES galaxies are local ($z\approx 0$).

\subsubsection{Distance}

Distance is measured through a variety of methods in PROBES, the most common being Hubble-Lema\^{i}tre flow distances, but it also includes, in a few cases, surface brightness fluctuations, the tip of the red giant branch, variable stars, cluster distances, and supernovae light-curve distance measurements.
Redshift-independent distances are mostly used in the SPARC and \citet[][hereafter SHIVir]{Ouellette2017} compilations. 
Secondary distance indicators based on galaxy scaling relations, such as the TFR, are left out as they form the basis of the present analysis.

Distance uncertainties are conservatively assumed to be \wunits{15}{\%} for Hubble-Lema\^{i}tre flow distances; 
if more accurate distance estimates are available in the original survey, those values are used instead.
A subset of the galaxies in \citet{Dale1999} are missing distance measurements; these and their associated uncertainty values are supplemented using NED distance measurements and are required for several parameters in \Sec{extractedparameters}, meaning that the distance uncertainties are correlated for scaling relations involving these parameters. 
The ability to effortlessly account for these correlated uncertainties is a major strength of the Bayesian intrinsic scatter method.

\subsubsection{Morphological Type}
\label{sec:uncertaintymodelhubbletype}

Morphological, or Hubble, types are provided for most original surveys in PROBES.
Any missing Hubble types were supplemented using NED.
No uncertainty is associated with Hubble types as they were mostly used for diagnostic purposes while developing our analysis, though they do enter into Model 1 of \Sec{inclinationcorrections}.
Hubble types are coded from 0 to 10 as: S0, Sa, Sab, Sb, Sbc, Sc, Scd, Sd, Sdm, Sm, and Im.

\subsection{Extracted Parameters}
\label{sec:extractedparameters}

\begin{figure*}[ht]
  \centering
  \includegraphics[width=0.8\textwidth]{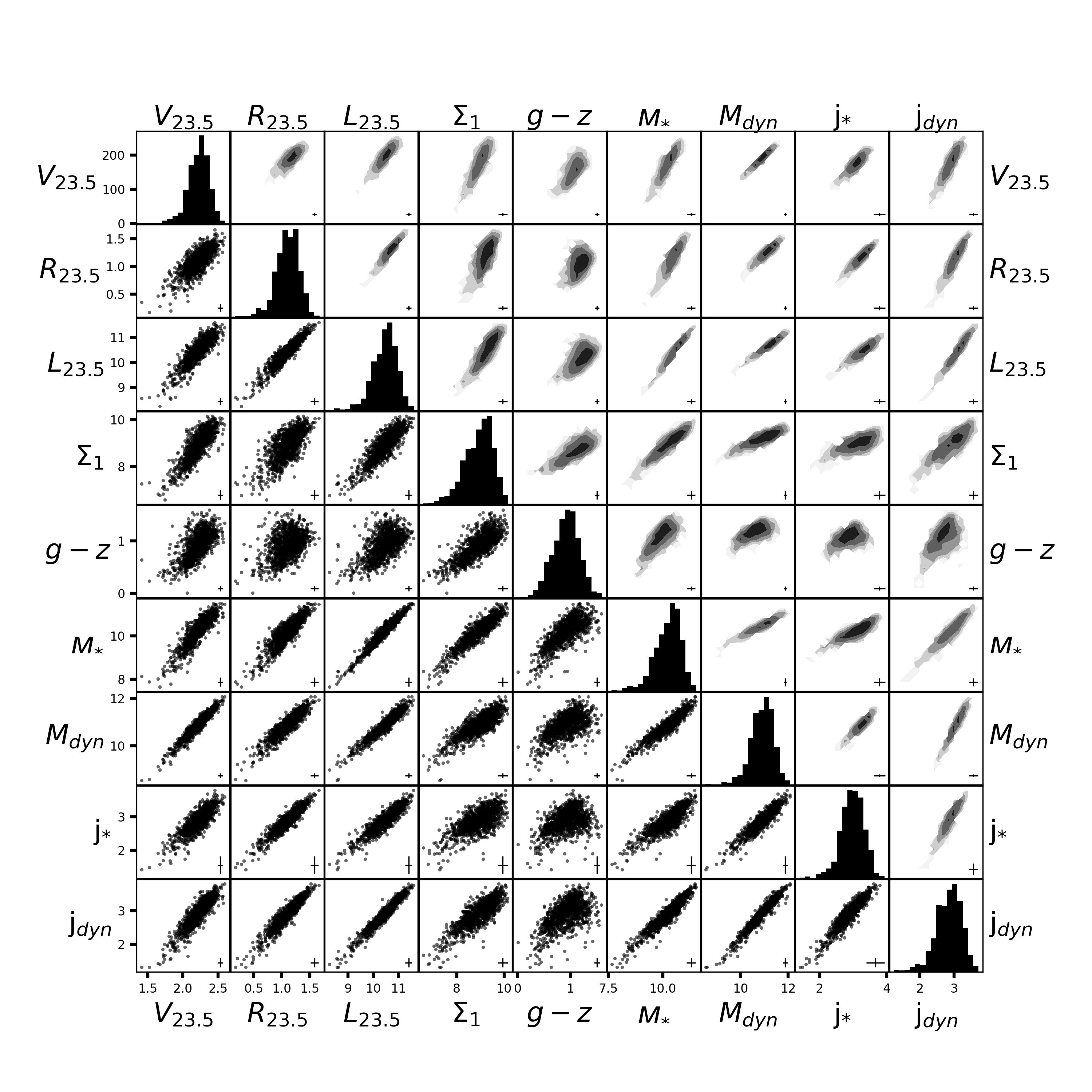}
  \caption{PROBES data with parameters in $\log$ space computed and corrected as described in this section (\Sec{datasets}).
  Every subplot is at the intersection of two variables. 
  In the lower triangle, scatter plots show each distribution. 
  Along the diagonal, histograms are shown for the given variable. 
  The upper triangle shows a density plot with four contours drawn evenly in log density. 
  The error bars in the bottom right corner of every non-diagonal subplot represent the median classical uncertainty (see \App{classicaluncertainty}).}
  \label{fig:TriangleParameters}
\end{figure*}

Once the data have been extracted from the raw observations, they can be processed into useful parameters.
Here, we present the nine core parameters used in our analysis.
In most descriptions, we include the largest sources of uncertainty, though these are not handled identically in the classical and Bayesian regimes.
The classical method described in \Sec{classicalintrinsicscatter} only works to first order and does not include covariances.  The equations for classical uncertainty can be found in \App{classicaluncertainty}.
The Bayesian method from \Sec{bayesianintrinsicscatter} propagates uncertainties while fully accounting for nonlinear functions and covariances between variables.
\Fig{TriangleParameters} shows the final results from this data extraction for all combinations of the studied parameters.
The layout of this figure will be used throughout the paper to present our analysis for all combinations of parameters in our multidimensional study.

\subsubsection{\texorpdfstring{${R_{23.5}~[kpc]}$}{R23.5 [kpc]}}
\label{sec:extractr23.5}

The galaxy size metric used in this analysis corresponds to the isophotal \wunits{23.5}{mag arcsec$^{-2}$} radius, converted to a physical radius using the distance for each galaxy.
All sizes in this paper are measured in the $z$ band as described in \Sec{desidata}.
This size metric yields minimal scatter in various scaling relations ~\citep{Hall2012,Trujillo2020}.
Incidentally, the radius $R_{23.5}$ measured at the $z$ band corresponds to a median stellar density of \wunits{6}{$M_{\odot}$/pc$^2$} with a $16-84$ quartile range of 
\wunits{4 - 10}{$M_{\odot}$/pc$^2$}.
The \wunits{1}{$M_{\odot}$/pc$^2$} radius advocated by \citet{Trujillo2020} thus samples radii larger than $R_{23.5}$.
Both size metrics produce an equally tight size - stellar mass relation~\citep{Trujillo2020Erratum}.
The range of stellar surface densities for $R_{23.5}$ also matches closely the predicted critical gas surface density for star formation in \citet{Schaye2004}.

$R_{23.5}$ is used as the metric at which all other parameters are measured, except $\Sigma_1$ which is measured at \wunits{1}{kpc} (\Sec{extractsigma1}). 
It is thus essential that the Bayesian intrinsic scatter model can account for correlated uncertainties, as many variables share $R_{23.5}$ at some point in their calculation.
$R_{23.5}$ is calculated as: 

\begin{equation}\label{equ:isophotalradius}
\begin{aligned}
    \log_{10}(R_{23.5}) &= \log_{10}(R_{\rm obs}D) +3 + C_{R_{23.5}},
\end{aligned}
\end{equation}

\noindent where $R_{\rm obs}$ is the \wunits{23.5}{mag arcsec$^{-2}$} isophotal radius in radians, taken directly from the surface brightness profile. 
$D$ is the distance in parsecs, and $C_{R_{23.5}}$ is the inclination correction factor (see \Sec{inclinationcorrections}).

\subsubsection{\texorpdfstring{${L_{23.5}~[L_{\odot}]}$}{L23.5 [Lsun]}}
\label{sec:extractL23.5}

The total luminosity in the $z$ band, $L_{{23.5}}$ integrated to $R_{23.5}$, is used as a standard metric for the total brightness of a galaxy. 
The luminosity is computed from a curve of growth, which is itself the integral of a surface brightness profile. 
Thus, all uncertainties related to zero-point calibrations and isophote uncertainties are carried forward.
$L_{{23.5}}$ is calculated as described in \Equ{luminosity} below:

\begin{equation}\label{equ:luminosity}
\begin{aligned}
    \log_{10}(L_{{23.5}}) &= \frac{(M_{\odot} - m_z)}{2.5} + 2\log_{10}\left(\frac{D}{10}\right) + C_{L_{23.5}},
\end{aligned}
\end{equation}

\noindent where $M_{\odot}$ is the solar absolute magnitude calibration, $m_z$ is the apparent magnitude in the $z$ band at $R_{23.5}$, and $C_{L_{23.5}}$ is the inclination correction factor (see \Sec{inclinationcorrections}).
We adopt $L_{23.5}$ as our preferred luminosity over other metrics, such as the last point in the light profile or extrapolating to infinity, both because it is evaluated at a consistent location in the galaxy profile and it encompasses most of the total light.
For comparison, the total light evaluated at the last point of the profile is on average (median) only \wunits{0.03}{dex} greater than $L_{23.5}$.

\subsubsection{\texorpdfstring{${V_{23.5}~[km~s^{-1}]}$}{V23.5 [km/s]}}
\label{sec:extractV23.5}

The rotation velocity of each galaxy is computed at the $R_{23.5}$ radius and corrected for inclination with the $1/\sin(i)$ factor and for redshift broadening with the $1/(1+z)$ factor, where $z$ is the heliocentric velocity of the galaxy divided by the speed of light.
Our measurement of $V_{23.5}$ uses a fit of the \citet{Courteau1997} multiparameter model to the observed rotation curve.
For rotation curves that do not extend to $R_{23.5}$, the fit is extrapolated out to the required radius. 
A substantial \wunits{62}{\%} of the PROBES rotation curves require extrapolated estimates for $V_{23.5}$, although over half of these extrapolations are less than \wunits{25}{\%} beyond the last point in the rotation curve.
Once extracted, the velocity is corrected for inclination and redshift broadening as indicated in \Equ{velocity}.

\begin{equation}\label{equ:velocity}
\begin{aligned}
    \log_{10}(V_{23.5}) &= \log_{10}\left(\frac{V_{\rm obs} - V_{\rm sys}}{\sin(i)(1+z)}\right), 
\end{aligned}
\end{equation}

\noindent where $V_{\rm obs}$ is the observed velocity computed from the \citet{Courteau1997} model, $V_{\rm sys}$ is the systematic velocity which is one of the model fit parameters, $i$ is the inclination in radians, and $z$ is the heliocentric redshift.
All PROBES galaxies are relatively nearby, and the respective redshift corrections (and their uncertainties) are quite small. 

\subsubsection{\texorpdfstring{${g-z~[mag]}$}{g-z [mag]}}
\label{sec:extractcolour}

Colours for PROBES galaxies are computed as the difference of the $g-$ and $z-$ band magnitudes measured at $R_{23.5}$ in the DESI-LIS photometry processed as described in \Sec{desidata}.
We choose $g-z$ colour over $g-r$ (which can also be measured from the DESI-LIS) for its longer baseline and low measurement error.
Colours are useful as inputs to colour mass-to-light transformations~\citep{Taylor2011,Conroy2013,Roediger2015,Zhang2017,Garcia-Benito2019} and can be used to construct scaling relations on their own.
The $g-z$ colours are calculated as:

\begin{equation}\label{equ:colour}
\begin{aligned}
    g-z &= m_g - m_z + C_{g-z},
\end{aligned}
\end{equation}

\noindent where $m_g$ is the apparent magnitude in the $g$ band, $m_z$ is the apparent magnitude in the $z$ band, and $C_{g-z}$ is the inclination correction factor. 
While the $g-$ and $z-$band magnitudes may be corrected independently first and then have both corrections added to the colour, we have found that the two corrections were highly correlated.
It is therefore more accurate in the classical framework to use a single correction factor. 

\subsubsection{\texorpdfstring{${M_*~[M_{\odot}]}$}{M* [Msun]}}
\label{sec:extractM*}

Stellar masses are calculated using the $z$ band luminosity (\Sec{extractL23.5}) and $g-z$ colour (\Sec{extractcolour}) input to the \citet{Roediger2015} colour to mass-to-light transformations (based on the \citet{Bruzual2003} stellar population model).
A random error of \wunits{0.05}{dex} is assumed for mass-to-light transformations~\citep[see Sec. 4.1 of][]{Roediger2015}.
Errors on colour/luminosity measurements are propagated separately.
For the mass range of most galaxies in PROBES, the \citet{Roediger2015} mass-to-light transformations are consistent with other transformations such as \citet{Zhang2017}.
Numerous uncertainty factors affect stellar mass estimates and are described in the luminosity and colour sections (\Secs{extractL23.5}{extractcolour}) 
\Equ{stellarmass} expresses the stellar mass calculation.

\begin{equation}\label{equ:stellarmass}
\begin{aligned}
    \log_{10}(M_{*}) &= \log_{10}(L_{23.5}\Upsilon(g-z)), 
\end{aligned}   
\end{equation}

\noindent where $\Upsilon(g-z)$ is the mass-to-light ratio as a function of the colour $g-z$. 
The luminosity and colour are taken as the corrected values in this formula, and no further inclination correction is needed. 
The stellar mass is, by construction, highly correlated with the luminosity and colour; this is challenging to adequately represent in a classical framework.

\subsubsection{\texorpdfstring{${\Sigma_1~[\text{M}_{\odot}~\text{kpc}^{-2}]}$}{Sigma1 [Msun/kpc2]}}
\label{sec:extractsigma1}

The quantity $\Sigma_1$ refers to the stellar-mass surface density within a radius of \wunits{1}{kpc}~\citep{Cheung2012,Fang2013,Zolotov2015,Teimoorinia2016,Chen2020}.
Our $\Sigma_1$ values are computed using $z$ band luminosity (out to \wunits{1}{kpc}), $g-z$ colour (out to \wunits{1}{kpc}), and the \citet{Roediger2015} colour to mass-to-light transformations as detailed in \Sec{extractM*}.
\Equ{sigma1} details our stellar-mass surface density calculation.

\begin{equation}\label{equ:sigma1}
\begin{aligned}
    \log_{10}(\Sigma_1) &= \frac{(M_{\odot} - m_{1,z})}{2.5} + 2\log_{10}\left(\frac{D}{10}\right) \\ &+\log_{10}\left(\frac{\Upsilon((g-z)_1)}{\pi}\right) + C_{\Sigma_{1}},
\end{aligned}
\end{equation}

\noindent where $C_{\Sigma_{1}}$ is the inclination correction for $\Sigma_{1}$, $m_{1,z}$ is the magnitude within \wunits{1}{kpc}, and $(g-z)_1$ is the colour within \wunits{1}{kpc}. 
The determination of the physical radius at \wunits{1}{kpc} requires a distance measurement.  
Distance and mass-to-light transformation are the most significant sources of uncertainty in $\Sigma_1$. 
Because of their common dependence on distance, $\Sigma_1$ and $R_{23.5}$ are strongly correlated even though they are measured at two different fiducial radii.
Obtaining measurements within \wunits{1}{kpc} requires galaxies to be well resolved; nearly \wunits{95}{\%} of the PROBES galaxies have seeing lengths less than \wunits{1}{kpc}, and are therefore sufficiently resolved for the calculation of $\Sigma_1$.
We calculate $\Sigma_1$ for all galaxies even if \wunits{1}{kpc} is not in principle resolved; 
the unresolved \wunits{5}{\%} of the PROBES galaxies shows a small bias of the order of \wunits{0.1}{dex} in their residuals for some scaling relations. 

\subsubsection{\texorpdfstring{${M_{\rm dyn}~[M_{\odot}]}$}{Mdyn [Msun]}}
\label{sec:extractMtot}

We compute the total mass out to $R_{23.5}$ using the observed rotation curves corrected for inclination.
\Equ{totalmass} expresses the total mass calculation, which is simply a combination of results from \Equ{velocity} and \Equ{isophotalradius} using the virial theorem~\citep{Binney2008}.

\begin{equation}\label{equ:totalmass}
\begin{aligned}
    \log_{10}(M_{\rm dyn}) &= \log_{10}\left(\frac{V_{23.5}^2R_{23.5}}{G}\right),
\end{aligned}   
\end{equation}

\noindent where $G$ is the gravitational constant (all other quantities are as defined above). 
As would be expected, the total mass is highly correlated with velocity and radius; however, it is also correlated with other variables such as luminosity (and stellar mass) since they share distance as a large source of uncertainty. 
The Bayesian uncertainty method (\Sec{bayesianintrinsicscatter}) is especially well suited to handle strong correlations of this kind, such as those that arise in the context of the stellar mass-halo mass relation (\Sec{stellarmasstotalmass}). 

\subsubsection{\texorpdfstring{${j_{\rm dyn}~[kpc~km~s^{-1}]}$}{jdyn [kpc km/s]}}
\label{sec:specificangularmomentum}

The dynamical angular momentum of a galaxy is a conserved quantity (in isolation), making it an effective lens into the history of galaxy formation.
The specific angular momentum $j_{\rm dyn}=J_{\rm dyn}/M_{\rm dyn}$ removes the trivial correlation between mass and angular momentum, allowing the physical relationship between these parameters to be studied more directly.

The specific angular momentum, $j_{\rm dyn}$, is computed out to $R_{23.5}$ by integrating the velocity profile found by fitting the \citet{Courteau1997} multiparameter model. 
The use of a model, instead of interpolating between galaxy velocity measurements, is more robust to noise, though some angular momentum information may be lost.
Note that the largest deviations from the \citet{Courteau1997} model occur near the center of the galaxy where contributions to angular momentum are suppressed by a factor of $R$, so ultimately the model's favourable behaviour against noise produces more accurate angular momentum measures.

The expression for specific angular momentum is shown in \Equ{specificangularmomentum}: 

{\footnotesize
\begin{equation}\label{equ:specificangularmomentum}
\begin{aligned}
    \log_{10}(j_{\rm dyn}) &= \log_{10}\left(\int_{0}^{R_{23.5}}\frac{V(r)^3r}{G}dr\right) - \log_{10}(M_{\rm dyn}),
\end{aligned}
\end{equation}}

\noindent where $V(r)$ is the rotational velocity as a function of radius, corrected for inclination and redshift as in \Sec{extractV23.5}.
$R_{23.5}$ and $M_{\rm dyn}$ are as described in \Secs{extractr23.5}{extractMtot}, respectively.
Note that we only use deprojected quantities in this formula and no further inclination correction is required.
Because the angular momentum profiles do not level off at $R_{23.5}$, our profiles are sensitive to the choice of cutoff radius.
Therefore, our results describe the angular momentum within $R_{23.5}$ and not the total angular momentum.

\subsubsection{\texorpdfstring{${j_{*}~[kpc~km~s^{-1}]}$}{j* [kpc km/s]}}

The stellar specific angular momentum is computed by integrating the stellar surface density and velocity out to $R_{23.5}$.
As in \Sec{specificangularmomentum}, the \citet{Courteau1997} multiparameter model is used to infer the velocity component.
Thanks to the fine sampling and high signal-to-noise of our photometry, we can forgo any model fitting and use the interpolated stellar surface density profile ($I(r)\Upsilon(r)$) directly from our isophotal fitting solution.
The stellar angular momentum in our analysis is calculated as in \Equ{stellarangularmomentum} as follows:

{\footnotesize
\begin{equation}\label{equ:stellarangularmomentum}
\begin{aligned}
    \log_{10}(j_{*}) &= \log_{10}\left(\frac{2\pi}{M_*}\int_{0}^{R_{23.5}}I(r)\Upsilon(r)V(r)r^2dr\right) + C_{J_{*}},
\end{aligned}
\end{equation}}

\noindent where $I(r)$ is the intensity as a function of radius, and $\Upsilon(r)$ is the mass-to-light ratio as a function of radius (we use the same transformations as in \Sec{extractM*}). 
$C_{J_{*}}$ is the inclination correction factor. 
The $R_{23.5}$ term in the integral is the corrected quantity from \Sec{extractr23.5} and the correction factor $C_{J_{*}}$ is assumed to only correct for the inclination dependence of the intensity profile $I(r)$ and the mass-to-light profile $\Upsilon(r)$.
As with $j_{\rm dyn}$, the stellar angular momentum profiles do not converge within $R_{23.5}$ and thus do not represent the total stellar angular momentum.

\subsection{Inclination Corrections}
\label{sec:inclinationcorrections}

Most galaxy structural parameters, detailed in \Sec{extractedparameters}, show some degree of correlation with inclination.
Any parameter correlation with inclination is undesirable as fundamental galaxy properties should not depend on their apparent orientation and projection on the sky. 
In the case of $V_{23.5}$, this dependence can be counteracted by dividing by $\sin(i)$; for other parameters, the transformation is not so trivial.

There are three primary sources of inclination dependence: (i) projection (geometric effect) on the sky; (ii) radiative transfer through the distribution of dust and gas in a galaxy, and (iii) stellar population distributions (or really any other spatially varying quantity). 
Projection effects can often be handled analytically, for example, in the case of surface brightness \citep{Byun1994}.
The brightness (or intensity per unit area) of an ideal galaxy taken as a transparent infinite slab will increase with inclination relative to face-on orientation since it is being projected through a larger length of the slab.
The increased brightness could be corrected by multiplying the linear flux by $\cos(i)$ in order to recover the face-on value from any inclination \citep{Holmberg1958, Giovanelli1994}.
However, galaxies are not ideal transparent slabs; instead, they have complex dust, gas, and stellar structures whose nature and distribution may vary from galaxy to galaxy.
We have attempted to correct surface brightnesses (and thus all subsequent extracted parameters) with a pure geometric correction of $-2.5\log_{10}(\cos(i))$.  
However, this yielded larger scatter estimates for many of our scaling relations.
Clearly, more than simple geometric corrections are at play.
A more generalized transformation is required to model these effects.

\begin{center}
\begin{deluxetable*}{c l c}
\tabletypesize{\footnotesize}
\tablewidth{\textwidth}
\tablecaption{Inclination Correction Models\label{tab:inclinationcorrectionmodels}}
\tablehead{Index & Model & Inspiration \\
(1) & (2) & (3)} 
\tablecolumns{3}
\startdata
        0 & $C(i) = \alpha_0 + \gamma_0\log_{10}(\cos(i))$ & \citet{Giovanelli1994} \\
        1 & $C(i,T) = \alpha_{[T]} + \gamma_{[T]}\log_{10}(\cos(i))$ & \citet{Han1992} \\
        2 & $C(i,V_{23.5}) = \alpha_1 + (\gamma_1 + \gamma_2\log_{10}(V_{23.5}))\log_{10}(\cos(i)) + \alpha_2\log_{10}(V_{23.5})$ & \citet{Tully1998} \\
        3 & $C(i,V_{23.5},C_{28}) = \alpha_3 + (\gamma_3 + \gamma_4\log_{10}(V_{23.5}) + \gamma_5C_{28})\log_{10}(\cos(i)) + \alpha_3\log_{10}(V_{23.5}) + \alpha_4C_{28}$ & \citet{Maller2009} \\
\enddata
\tablecomments{Column (1) indexes the inclination correction models. Column (2) gives the model. Column (3) lists the literature source that inspired the correction model. Note that our model may differ from the quoted source, but is conceptually related (see the text for more details). A subscript $T$ means that the parameter changes with morphological type.}
\end{deluxetable*}
\end{center}

The ideal form of a generalized inclination correction model is unknown, though many variations have been tried since at least \citet{Holmberg1958} who used a cosec function.
The most common approach takes the form detailed in \citet{Giovanelli1994} and included as Model 0 in \Tab{inclinationcorrectionmodels}~\citep{Burstein1991,Han1992,Willick1995,Tully1998,Shao2007,Graham2008,Unterborn2008,Cho2009,Maller2009,Masters2010,Devour2016,Devour2017,Devour2019}. 
Other forms have been explored as well~\citep{Mollenhoff2006,Driver2007,Driver2008,Shao2007,Tempel2010,Yip2010,Xiao2012,Kourkchi2019}.
The two common parameterizations of inclination dependence rely on a function of either the axial ratio, $b/a$, or the inclination $i$ (from \Equ{inclination}). 
The latter accounts for the thickness of the disk via a (poorly constrained) flattening parameter $q_0=c/a$, while the former does not. 
Dust-free images of edge-on galaxies show that the values $q_0$ are clearly nonzero and so our modest assumed value ($q_{0} = 0.13$, \citealt{Hall2012}) should provide a closer approximation to the truth. 
The factors $\alpha_i,\gamma_i$ in \Tab{inclinationcorrectionmodels} encapsulate our ignorance about the exact nature of the obscuration in a galaxy as a function of inclination.
The nature of obscuration may vary with disk thickness, dust distribution, disk features, and other unknown factors in a galaxy.
As such, the coefficients may change for different galaxy types.  
\citet{Han1992} divided their sample into Hubble type bins before computing their correction factor; however, later studies have used other techniques.
\citet{Tully1998} divided galaxies into magnitude bins then used the TFR to reformat the correction factor $\gamma$ as a function of velocity
(see also \citet{Willick1997}).
\citet{Driver2007,Driver2008} and \citet{Masters2010} separated galaxies into ``bulgy'' and ``disky'' groups in order to compute their $\gamma$.
\citet{Maller2009} tried two subdivision techniques, one based on magnitude and the other based on S\'ersic index, while \citet{Cho2009}'s corrections used concentration index as a representation of morphology.
In their related series of papers, \citet{Devour2016, Devour2017,Devour2019} exploited a two-dimensional space of magnitude and colour in infrared bands.
The above works establish a consensus that inclination corrections depend on some notion of galaxy ``families'' with similar properties.
However, the range of techniques speaks to the difficulty of finding parameters that are not themselves inclination dependent.

To generalize the concept of inclination correction, we consider a general model $C$ that expresses an extracted parameters dependence on inclination.
The model may depend on any galactic parameter, in principle, though we restrict our consideration to inclination, velocity, morphological type, and concentration as these are (mostly) independent of inclination themselves.
We can then determine an inclination correction by fitting $X = C(i,V_{23.5},T,C_{28})$ where $X$ is the extracted parameter (\Sec{extractedparameters}), $i$ is inclination, $V_{23.5}$ is the rotation velocity, $T$ is morphological type, and $C_{28}$ is the light concentration index.
The fit is performed with a least-squares regression, then the inclination correction is taken to be: $C(0,V_{23.5},T,C_{28}) - C(i,V_{23.5},T,C_{28})$ which corrects to a face-on value\footnote{Internal extinction is still present in face-on systems. Its correction would require a wavelength-dependent radiative transfer code. Such a treatment is beyond the scope of the present study.}.
Any term in the model that does not depend on inclination will ultimately be absent from the correction and is only used for the sake of fitting.
For clarity, the coefficients for these ``absent'' terms are labeled with $\alpha$.

\begin{center}
\begin{deluxetable}{c c c c c c c}
\tabletypesize{\footnotesize}
\tablewidth{\columnwidth}
\tablecaption{Inclination Correction Coefficients\label{tab:inclinationcorrectioncoefficients}}
\tablehead{Model & Coefficient & $L_{23.5}$ & $g-z$ & $R_{23.5}$ & $\Sigma_{1}$ & $j_{*}$ \\
    (1) & (2) & (3) & (4) & (5) & (6) & (7)} 
\tablecolumns{7}
\startdata
    0 & $\alpha_0$ & 10.40 & 0.91 & 1.08 & 8.87 & 2.91 \\ 
    0 & $\gamma_0$ & 0.33 & -0.49 & -0.15 & 0.06 & -0.72 \\ 
    1 & $\alpha_{[0-3]}$ & 10.46 & 0.97 & 1.09 & 9.06 & 2.91 \\ 
    1 & $\gamma_{[0-3]}$ & -0.10 & -0.58 & -0.31 & -0.30 & -0.96 \\ 
    1 & $\alpha_{[4]}$ & 10.61 & 0.92 & 1.17 & 9.05 & 3.04 \\ 
    1 & $\gamma_{[4]}$ & 0.48 & -0.51 & -0.07 & 0.13 & -0.63 \\ 
    1 & $\alpha_{[5]}$ & 10.39 & 0.89 & 1.09 & 8.72 & 2.93 \\ 
    1 & $\gamma_{[5]}$ & 0.59 & -0.51 & -0.02 & 0.10 & -0.51 \\ 
    1 & $\alpha_{[6]}$ & 10.45 & 0.88 & 1.16 & 8.67 & 3.06 \\ 
    1 & $\gamma_{[6]}$ & 0.89 & -0.26 & 0.16 & 0.35 & -0.25 \\ 
    1 & $\alpha_{[7-10]}$ & 9.91 & 0.79 & 0.84 & 8.48 & 2.56 \\ 
    1 & $\gamma_{[7-10]}$ & 0.08 & -0.27 & -0.41 & 0.44 & -1.12 \\ 
    2 & $\alpha_1$ & 4.29 & -0.65 & -1.34 & 2.63 & -1.59 \\ 
    2 & $\gamma_1$ & 1.48 & 1.54 & -0.02 & 3.24 & -1.86 \\ 
    2 & $\gamma_2$ & -0.58 & -0.95 & -0.08 & -1.51 & 0.49 \\ 
    3 & $\alpha_2$ & 4.29 & -0.65 & -1.34 & 2.63 & -1.59 \\ 
    3 & $\alpha_3$ & 1.58 & -0.43 & 0.69 & -0.04 & 0.40 \\ 
    3 & $\alpha_4$ & 1.18 & 1.13 & 0.40 & 2.84 & 1.63 \\ 
    3 & $\gamma_3$ & 1.48 & 1.54 & -0.02 & 3.24 & -1.86 \\ 
    3 & $\gamma_4$ & -0.89 & -0.44 & -0.39 & 0.55 & 0.61 \\ 
    3 & $\gamma_5$ & 0.31 & -0.50 & 0.31 & -2.06 & -0.12 \\ \hline
\enddata
\tablecomments{Fitted coefficients for all models in \Tab{inclinationcorrectionmodels}. Column (1) is the model index. Column (2) lists the coefficient from \Tab{inclinationcorrectionmodels} being fit. Columns (3) - (7) give the coefficient values for each corrected parameter.}
\end{deluxetable}
\end{center}

We considered several models to describe inclination correlations; four such models are listed in \Tab{inclinationcorrectionmodels} for comparison.
Model 0 is a base case where all galaxies are considered, without regard to subdivision by ``family.'' 
Model 1 divides galaxies into bins by morphological type and fits the coefficients $\alpha_T,\gamma_T$ for each bin.
This represents the simplest notion of inclination correction family.
Model 2 allows the coefficient on $\log_{10}(\cos(i))$ to be a function of velocity (now expressed as $\gamma_1,\gamma_2$).
Note that \citet{Tully1998} fit coefficients to the residuals of a colour-magnitude fit, while we fit the coefficients directly to the extracted parameter.
\citet{Tully1998} used the color $B-K'$ and the dependence of $K'$ on inclination was assumed to be negligible.
Model 4 is inspired by \citet{Maller2009}, who used infrared wavelength photometry and S\'ersic indices.
However, the TFR indicates that magnitude in any band (and especially infrared bands) is a tight function of velocity, and so we use it in its stead.
Similarly, instead of the S\'ersic index, we prefer using the model-independent light concentration index $C_{28}$.
\citet{Maller2009} fitted their data with their Equation~6 for nearly face-on galaxies, and then fitted the residuals with their $\alpha$ coefficients from Equation~9 using a $\chi^2$ minimization.
Our Model 4 encodes this two-step residual fitting into a single operation, though it gives no explicit preference to near face-on galaxies at any point.
\Tab{inclinationcorrectioncoefficients} lists the coefficients fitted to the models in \Tab{inclinationcorrectionmodels}.

\begin{figure}
    \centering
    \includegraphics[width=\columnwidth]{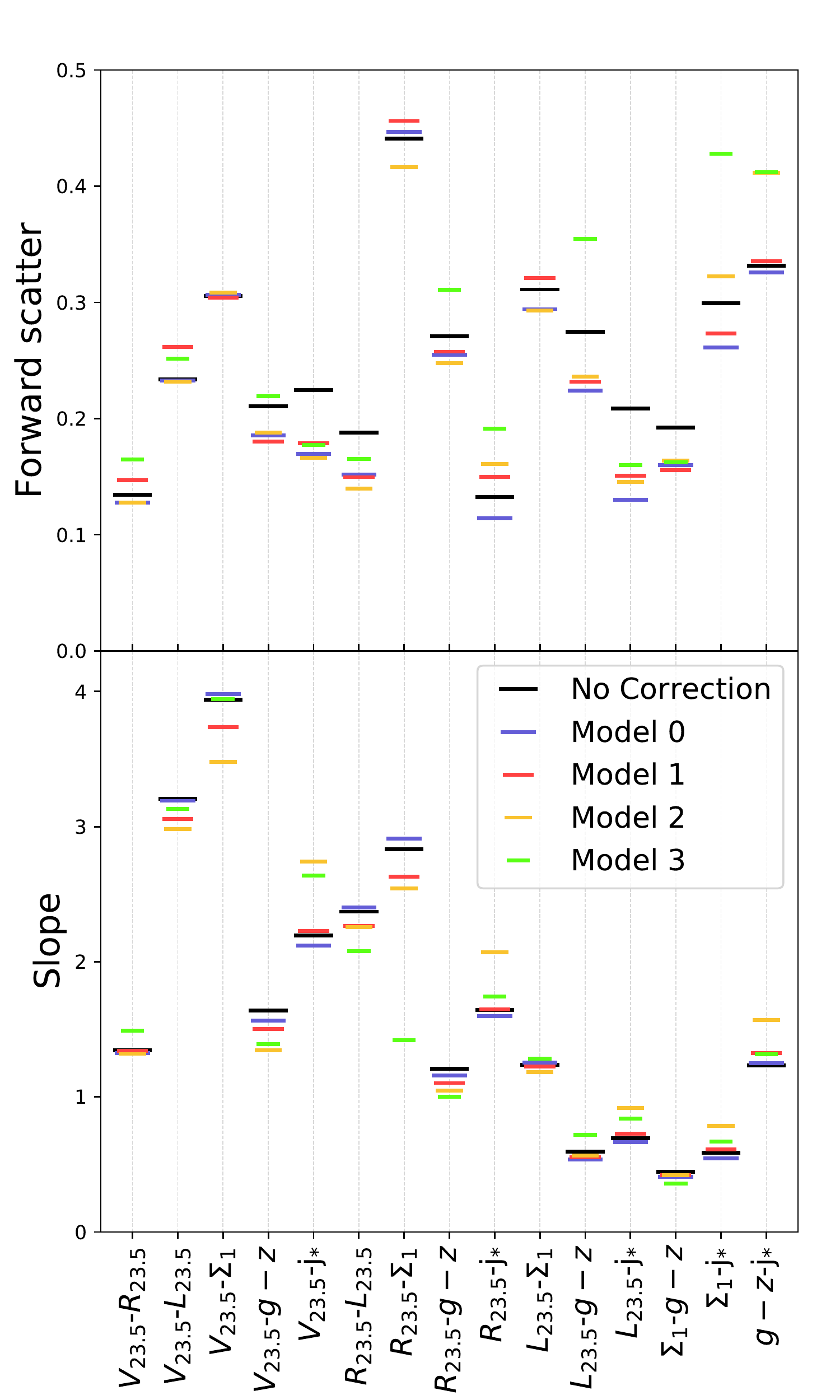}
    \caption{Comparison of scatters and slopes for inclination correction models from \Tab{inclinationcorrectionmodels}. The scatter (top) and slope (bottom) on the $y$-axis correspond to the scaling relations on the $x$-axis. 
    For clarity, only scaling relations including inclination-corrected quantities are shown, as well as velocity. Most scatters are in dex, though any relations with $g-z$ as the $y$-variable is in units of magnitude. Model 3 is not shown for relations where the scatter/slope is too large.}
    \label{fig:correctioncompare}
\end{figure}

After fitting the various models, we examined their effects on our suite of galaxy scaling relations.
\Fig{correctioncompare} shows the variation of the scatter and slope with each correction model.
Only relations involving parameters that were directly inclination-corrected, and velocity, are included for clarity.
Relations not included in the figure showed similar variability with the inclination correction model.
Applied to our data, these techniques had mixed effects on our scaling relations, alternately increasing and decreasing the slope/scatter of many scaling relations.
Ideally, an accurate model should yield a broad scatter reduction for an ensemble of scaling relations.
The only model that fits this description is Model 0, which has no notion of galaxy ``families'' despite the broad range of PROBES galactic properties.
Thus our analysis will use Model 0 for projection correction throughout, even though the true correction likely depends on some notion of galaxy family.
Our data are likely not sensitive enough to reveal the nature of (putative) galaxy families.

\Fig{correctioncompare} also reveals the challenging nature of inclination corrections and the power of examining many scaling relations simultaneously.
This figure shows that any inclination correction model does confer a reduced scatter in some cases and an increased scatter in others.
An inclination correction scheme based on the examination of a single scaling relation could easily map to non-intrinsic quantities.

\subsection{Bayesian Error Budget}

\begin{figure*}[ht]
  \centering
  \includegraphics[width=0.9\textwidth]{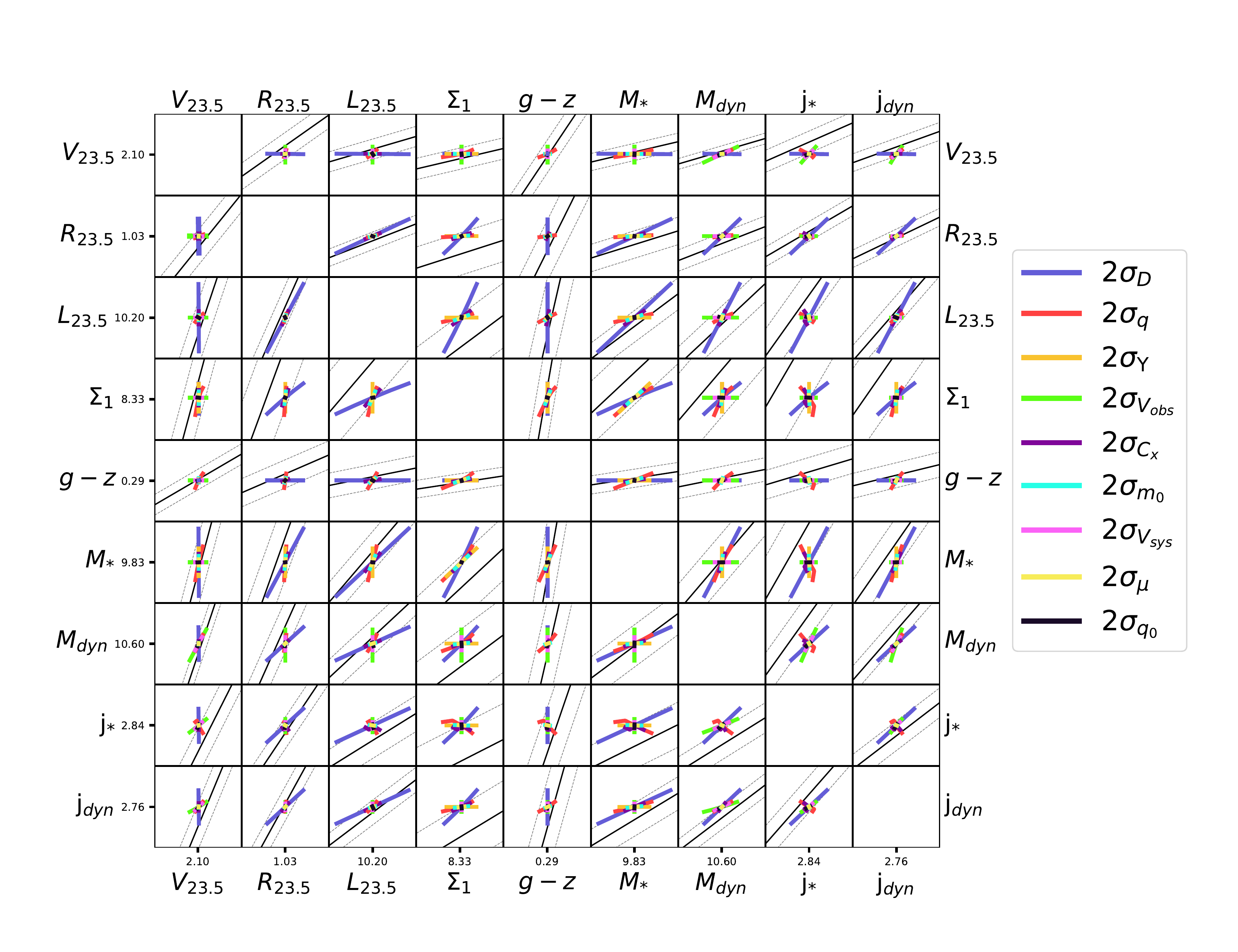}
  \caption{
 Representation of major sources of uncertainty in galaxy scaling relations. 
 This demonstration applies to the typical galaxy, UGC~12521 \citep{Courteau1997}.
 Formatted similarly to \Fig{TriangleParameters}, the intersection of any two variables gives the corresponding scaling relation. 
 The black solid line is the scaling relation fit as presented in \Tab{scalingrelationfits}, while the black dashed lines give the $1\sigma$ forward scatter from \Tab{scalingrelationscatters}.
 The legend indicates different sources of uncertainty, namely the distance, $D$; axial ratio, $q = b/a$; stellar mass-to-light ratio, $\Upsilon = M_*/L$; observed rotational velocity, $V_{\rm obs}$; projection correction, $C_x$ (all corrections considered simultaneously); photometric zero-point, $m_0$; recessional velocity, $V_{\rm sys}$; surface brightness, $\mu$; and disk flattening parameter, $q_0 = c/a$. 
 Coloured bars represent $2\sigma$ perturbations for every variable shown in the legend.
 Each variable is adjusted one at a time to show its effect in isolation.
 All subplots represent a window \wunits{0.6}{dex} across; the $g-z$ parameter is converted to dex for comparison with other parameters.
 Note that the specific features in each relation are expected to change for every galaxy.}
  \label{fig:Perturbations}
\end{figure*}

In the Bayesian formalism, all measurements with uncertainties are randomly sampled and refit with the above analysis methods.
As a result, a simplistic uncertainty propagation as shown in \App{classicaluncertainty} is not possible; instead, the movement of a point can only be traced when a given variable is perturbed.
\Fig{Perturbations} shows the effect of every source of uncertainty in our analysis for a single galaxy.
A similar figure could be produced for every galaxy and there would be some variability in the figures as sources of error change in relative significance; however, UGC~12521 exhibits many typical features.

Since the Bayesian method works by resampling all measurements and then recomputing all variables, some variables did not neatly map into \Fig{Perturbations}.
Velocity is determined by fitting a rotation curve comprising many measurements that are typically all individually resampled. 
To produce a clear signal, we instead increased/decreased all of the velocities by $2\sigma$.
This is unlikely to occur in a true random sampling scenario; the velocity uncertainty in the figure is thus a worst-case scenario.
The surface brightness profile data were also adjusted in the same direction by $2\sigma$ causing the whole profile to shift (in proportion to the surface brightness, SB, uncertainty), again making this a worst-case scenario for the amount of shift in \Fig{Perturbations}.
To account for the systematic velocity uncertainty, the profiles were adjusted after fitting the \citet{Courteau1997} multiparameter model. 
A systemic velocity shift prior to fitting the model would be canceled by the fitting routine.

\Fig{Perturbations} shows that all scaling relations have some correlated uncertainties, as well as the relative scatter dependencies for each relation. 
Distance is typically the dominant source of error. 
Photometric zero-point, mass-to-light ratio, velocity, and axial ratio are also significant contributors. 
Curvature can also be seen in some scaling relations (i.e. the $j_*$ versus $g-z$ relation);  
the proper error propagation for those can only be captured in the Bayesian analysis.

Anticorrelated errors, such as those in most relations involving $j_*$ and the axis ratio $q$, contribute to a scatter increase in a manner that cannot be ascertained by classical analysis. 
The fits in each panel also show that correlated distance errors (and other correlated sources of error) are often closely, but not entirely, aligned with a scaling relation. 
Because correlated error vectors would not perfectly slide along the relation, a classical analysis would under(over)estimate intrinsic scatter if distance errors are included (excluded). 

\subsection{Data Quality Cuts}
\label{sec:dataqualitycuts}

Various data cuts were applied for quality control. 
Tests in \App{testtruncation} suggest that quality cuts based on parameter uncertainty do not bias intrinsic scatter measurements, while cuts based on deviations from a relation (often referred to as ``sigma clipping'') drastically bias scatter measurements. 
Our fitting method (described in \Sec{scalingrelationfits}) is also robust to quality cuts; 
note that sigma clipping is never used in our analysis, even as part of our fitting procedure.
Further tests on our fitting routine and intrinsic scatter estimators were performed for a variety of data modifications.
The fitting algorithm is robust to realistic sampling biases (e.g. magnitude limited data), covariant observational uncertainties, and slightly incorrect estimates of observational uncertainty.
See \App{testtruncation} for more details.

\Tab{dataqualitycuts} lists the cuts for the removal of a galaxy from our analysis.
Multiple cuts may remove the same galaxy so the totals given for each cut cannot be simply added together.
Note as well that some cuts may not remove galaxies from the analysis; however, they may be used for the Bayesian intrinsic scatter calculations if the random sampling generates a pathological galaxy sample.

Starting with 1396 galaxies, we apply our data quality cuts to give 1152 high-quality galaxy samples.
Note that we apply our Bayesian analysis to the full original sample, and simulate the effect of our data quality cuts.
This means that some galaxies may enter into our Bayesian intrinsic scatter calculation if they are close to the edge of of a data quality cut.

\begin{center}
\begin{deluxetable}{c c c c c c}
\tabletypesize{\normalsize}
\tablewidth{\columnwidth}
\tablecaption{Data Quality Cuts\label{tab:dataqualitycuts}}
\tablehead{Variable & Units & $X \geq$ & $X \leq$ & $\sigma \leq$ & N-removed \\ 
            (1) & (2) & (3) & (4) & (5) & (6)} 
\tablecolumns{6}
\startdata
$D$ & dex & - & - & 0.1 & 65 \\
$i$ & deg & 30 & 80 & 10 & 118 \\
$R_{23.5}$ & dex & -1 & 2 & 0.15 & 23 \\
$L_{23.5}$ & dex & 6.5 & 12 & 0.3 & 28 \\
$V_{23.5}$ & dex & 0.8 & 3 & 0.1 & 53 \\
$g-z$ & mag & -0.5 & 2 & 0.2 & 37 \\
$M_{*}$ & dex & 7 & 12 & 0.4 & 34 \\
$\Sigma_{1}$ & dex & 6 & 10.5 & 0.4 & 22 \\
$M_{\rm dyn}$ & dex & 6 & 13 & 0.2 & 76 \\
$j_{*}$ & dex & 0 & 4.5 & 0.4 & 61 \\
$j_{\rm dyn}$ & dex & 1 & 4 & 0.3 & 82 \\
\enddata
\tablecomments{Variables in Column (1) are described in \Secs{extractedparameters}{uncertaintymodel}.  
Column (2) indicates the units of the (log/linear) limits.  
Columns (3) and (4) give a lower and upper bound for each variable.
Column (5) gives an upper bound on uncertainty. 
Column (6) reports the number of galaxies cut from PROBES by limits on that variable, though some galaxies may be removed on account of multiple constraints.
No value/limit is set for the distance measurements.}
\end{deluxetable}
\end{center}

\section{Results}
\label{sec:results}

\subsection{Scaling Relation Fits}
\label{sec:scalingrelationfits}

Before evaluating the scatter around a scaling relation, we must first fit linear relations to all parameter combinations. 
However, the choice of regression model can significantly impact the resulting fitted parameters. 
For the study of galaxy scaling relations, both axes have heteroscedastic uncertainty (often covariant uncertainty) and neither can be considered an independent variable.
Furthermore, a nonzero intrinsic scatter is expected to be present in each scaling relation.

\begin{center}
\begin{deluxetable*}{c c c c c c c c c c c}
\tabletypesize{\footnotesize}
\tablewidth{\textwidth}
\tablecaption{Scaling Relation Fits\label{tab:scalingrelationfits}}
\tablehead{\colhead{$Y\diagdown X$} & \colhead{~} & \colhead{$V_{23.5}$} & \colhead{$R_{23.5}$} & \colhead{$L_{23.5}$} & \colhead{$\Sigma_{1}$} & \colhead{$g-z$} & \colhead{$M_{*}$} & \colhead{$M_{\rm dyn}$} & \colhead{$j_{*}$} & \colhead{$j_{\rm dyn}$}} 
\startdata
    \multirow{2}{*}{$V_{23.5}$} & $m$ & \multirow{2}{*}{$\bullet$} & $0.755_{0.040}^{0.019}$ & $0.313_{0.012}^{0.004}$ & $0.251_{0.002}^{0.004}$ & $0.640_{0.015}^{0.004}$ & $0.250_{0.009}^{0.002}$ & $0.314_{0.004}^{0.001}$ & $0.472_{0.019}^{0.004}$ & $0.386_{0.008}^{0.003}$\\
    ~  & $b$ & ~ & $1.385_{0.016}^{0.050}$ & $\llap{-}1.059_{0.041}^{0.128}$ & $\llap{-}0.025_{0.036}^{0.021}$ & $1.628_{0.009}^{0.021}$ & $\llap{-}0.341_{0.024}^{0.091}$ & $\llap{-}1.203_{0.002}^{0.043}$ & $0.850_{0.007}^{0.072}$ & $1.088_{0.017}^{0.031}$\\
    \hline
    \multirow{2}{*}{$R_{23.5}$} & $m$ & $1.325_{0.034}^{0.074}$ & \multirow{2}{*}{$\bullet$} & $0.417_{0.004}^{0.008}$ & $0.343_{0.013}^{0.021}$ & $0.862_{0.013}^{0.036}$ & $0.334_{0.004}^{0.009}$ & $0.418_{0.013}^{0.020}$ & $0.626_{0.002}^{0.020}$ & $0.513_{0.012}^{0.026}$\\
    ~  & $b$ & $\llap{-}1.835_{0.168}^{0.074}$ & ~ & $\llap{-}3.252_{0.098}^{0.027}$ & $\llap{-}1.961_{0.197}^{0.111}$ & $0.309_{0.046}^{0.001}$ & $\llap{-}2.313_{0.101}^{0.028}$ & $\llap{-}3.458_{0.223}^{0.135}$ & $\llap{-}0.711_{0.078}^{0.013}$ & $\llap{-}0.398_{0.075}^{0.041}$\\
    \hline
    \multirow{2}{*}{$L_{23.5}$} & $m$ & $3.191_{0.042}^{0.130}$ & $2.401_{0.046}^{0.021}$ & \multirow{2}{*}{$\bullet$} & $0.797_{0.020}^{0.035}$ & $1.862_{0.012}^{0.054}$ & $0.799_{0.001}^{0.008}$ & $1.003_{0.016}^{0.032}$ & $1.503_{0.020}^{0.027}$ & $1.232_{0.012}^{0.046}$\\
    ~  & $b$ & $3.379_{0.298}^{0.086}$ & $7.807_{0.023}^{0.099}$ & ~ & $3.344_{0.328}^{0.148}$ & $8.738_{0.081}^{0.043}$ & $2.286_{0.081}^{0.002}$ & $\llap{-}0.486_{0.360}^{0.167}$ & $6.098_{0.162}^{0.149}$ & $6.852_{0.174}^{0.008}$\\
    \hline
    \multirow{2}{*}{$\Sigma_{1}$} & $m$ & $3.980_{0.058}^{0.027}$ & $2.912_{0.169}^{0.107}$ & $1.255_{0.053}^{0.030}$ & \multirow{2}{*}{$\bullet$} & $2.453_{0.064}^{0.010}$ & $1.004_{0.035}^{0.021}$ & $1.257_{0.023}^{0.002}$ & $1.833_{0.084}^{0.030}$ & $1.541_{0.036}^{0.003}$\\
    ~  & $b$ & $0.098_{0.083}^{0.138}$ & $5.711_{0.076}^{0.224}$ & $\llap{-}4.196_{0.290}^{0.565}$ & ~ & $6.662_{0.035}^{0.103}$ & $\llap{-}1.342_{0.188}^{0.384}$ & $\llap{-}4.782_{0.003}^{0.257}$ & $3.609_{0.008}^{0.321}$ & $4.419_{0.055}^{0.142}$\\
    \hline
    \multirow{2}{*}{$g-z$} & $m$ & $1.564_{0.009}^{0.038}$ & $1.160_{0.046}^{0.017}$ & $0.537_{0.015}^{0.003}$ & $0.408_{0.002}^{0.011}$ & \multirow{2}{*}{$\bullet$} & $0.411_{0.010}^{0.002}$ & $0.538_{0.006}^{0.010}$ & $0.801_{0.017}^{0.007}$ & $0.654_{0.006}^{0.012}$\\
    ~  & $b$ & $\llap{-}2.545_{0.088}^{0.025}$ & $\llap{-}0.358_{0.007}^{0.063}$ & $\llap{-}4.693_{0.040}^{0.163}$ & $\llap{-}2.716_{0.108}^{0.005}$ & ~ & $\llap{-}3.285_{0.019}^{0.110}$ & $\llap{-}4.945_{0.107}^{0.072}$ & $\llap{-}1.400_{0.052}^{0.086}$ & $\llap{-}0.991_{0.047}^{0.032}$\\
    \hline
    \multirow{2}{*}{$M_{*}$} & $m$ & $3.997_{0.039}^{0.144}$ & $2.994_{0.078}^{0.034}$ & $1.252_{0.012}^{0.001}$ & $0.996_{0.021}^{0.036}$ & $2.431_{0.009}^{0.062}$ & \multirow{2}{*}{$\bullet$} & $1.256_{0.014}^{0.034}$ & $1.878_{0.036}^{0.022}$ & $1.542_{0.007}^{0.049}$\\
    ~  & $b$ & $1.363_{0.330}^{0.081}$ & $6.924_{0.013}^{0.135}$ & $\llap{-}2.861_{0.001}^{0.128}$ & $1.337_{0.349}^{0.161}$ & $7.985_{0.090}^{0.036}$ & ~ & $\llap{-}3.467_{0.373}^{0.144}$ & $4.782_{0.173}^{0.216}$ & $5.717_{0.192}^{0.028}$\\
    \hline
    \multirow{2}{*}{$M_{\rm dyn}$} & $m$ & $3.188_{0.003}^{0.040}$ & $2.391_{0.111}^{0.070}$ & $0.997_{0.031}^{0.016}$ & $0.796_{0.001}^{0.015}$ & $1.858_{0.035}^{0.022}$ & $0.796_{0.021}^{0.009}$ & \multirow{2}{*}{$\bullet$} & $1.497_{0.052}^{0.010}$ & $1.228_{0.016}^{0.015}$\\
    ~  & $b$ & $3.834_{0.092}^{0.003}$ & $8.266_{0.066}^{0.132}$ & $0.484_{0.156}^{0.332}$ & $3.804_{0.142}^{0.006}$ & $9.190_{0.038}^{0.045}$ & $2.761_{0.079}^{0.224}$ & ~ & $6.562_{0.023}^{0.201}$ & $7.313_{0.057}^{0.058}$\\
    \hline
    \multirow{2}{*}{$j_{*}$} & $m$ & $2.121_{0.017}^{0.088}$ & $1.598_{0.049}^{0.004}$ & $0.666_{0.012}^{0.009}$ & $0.546_{0.009}^{0.026}$ & $1.249_{0.011}^{0.028}$ & $0.533_{0.006}^{0.011}$ & $0.668_{0.004}^{0.024}$ & \multirow{2}{*}{$\bullet$} & $0.820_{0.010}^{0.023}$\\
    ~  & $b$ & $\llap{-}1.802_{0.228}^{0.010}$ & $1.136_{0.030}^{0.093}$ & $\llap{-}4.058_{0.141}^{0.167}$ & $\llap{-}1.969_{0.274}^{0.040}$ & $1.748_{0.081}^{0.061}$ & $\llap{-}2.547_{0.156}^{0.117}$ & $\llap{-}4.382_{0.287}^{0.025}$ & ~ & $0.502_{0.113}^{0.010}$\\
    \hline
    \multirow{2}{*}{$j_{\rm dyn}$} & $m$ & $2.590_{0.020}^{0.054}$ & $1.948_{0.096}^{0.046}$ & $0.811_{0.029}^{0.008}$ & $0.649_{0.001}^{0.016}$ & $1.528_{0.027}^{0.015}$ & $0.648_{0.020}^{0.003}$ & $0.814_{0.010}^{0.011}$ & $1.219_{0.034}^{0.015}$ & \multirow{2}{*}{$\bullet$}\\
    ~  & $b$ & $\llap{-}2.819_{0.139}^{0.063}$ & $0.774_{0.059}^{0.101}$ & $\llap{-}5.560_{0.051}^{0.333}$ & $\llap{-}2.867_{0.160}^{0.040}$ & $1.514_{0.040}^{0.048}$ & $\llap{-}3.707_{0.003}^{0.229}$ & $\llap{-}5.954_{0.126}^{0.116}$ & $\llap{-}0.612_{0.004}^{0.151}$ & ~\\
\enddata
\tablecomments{This table is formatted like \Fig{TriangleParameters} with results for each relation at the intersection of two variables. The first column gives the $y$-axis parameter for each scaling relation.  The second column gives the linear fit parameters where the fit is of the form: $Y = mX+b$. Reading across a row gives every parameter option as the $x$-variable. The diagonal cells are left blank. Parameter uncertainties are determined by Monte Carlo sampling.}
\end{deluxetable*}
\end{center}

Fits are performed using the BCES bisector algorithm of \citet{Akritas1996}.
These are presented in \Tab{scalingrelationfits} which is organized like \Fig{TriangleParameters}.
The BCES bisector algorithm can model covariant heteroscedastic uncertainties, which we extract from our Bayesian error analysis.
All scaling relations appear in the table twice to account for forward and inverse relations, though the fits are performed with a BCES bisector and so the forward and inverse fits are compatible.
The first column and first row are analogous to axes of a plot, and so for any value in the table, one can look at the column header to get the $x$-axis and the row variable to get the $y$-axis.
The uncertainties presented here include only random errors computed from the Bayesian method and so correspond to the posterior.  
They do not account for the systematic errors that result from galaxy sample selection or the choice of model.
As will be seen in \Sec{discussion}, systematic errors are the dominant source of discrepancies when scaling relation fits from independent studies are inter-compared.

The resulting fits include a large number of known scaling relations. 
These include relations such as stellar mass - luminosity that are too strongly correlated to be used for testing galaxy formation models.
While our values agree well with the literature (see \Sec{discussion}), there is a great variety in intercepts due to parameter choices making their comparison more challenging. 

\subsection{Intrinsic Scatter of Scaling Relations}
\label{sec:intrinsicscatterresults}

\begin{center}
\begin{deluxetable*}{c c c c c c c c c c c}
\tabletypesize{\footnotesize}
\tablewidth{\textwidth}
\tablecaption{Scaling Relation Scatters\label{tab:scalingrelationscatters}}
\tablehead{\colhead{$Y\diagdown X$} & \colhead{~} & \colhead{$V_{23.5}$} & \colhead{$R_{23.5}$} & \colhead{$L_{23.5}$} & \colhead{$\Sigma_{1}$} & \colhead{$g-z$} & \colhead{$M_{*}$} & \colhead{$M_{\rm dyn}$} & \colhead{j$_{*}$} & \colhead{j$_{\rm dyn}$}} 
\startdata
    ~  & $\sigma_o$ & ~ & $0.096_{0.003}^{0.003}$ & $0.073_{0.002}^{0.002}$ & $0.077_{0.002}^{0.002}$ & $0.119_{0.003}^{0.004}$ & $0.062_{0.002}^{0.002}$ & $0.039_{0.001}^{0.001}$ & $0.080_{0.003}^{0.002}$ & $0.065_{0.002}^{0.002}$\\
    $V_{23.5}$ & $\sigma_b$ & $\bullet$ & $0.086_{0.002}^{0.003}$ & $0.056_{0.002}^{0.002}$ & $0.073_{0.002}^{0.002}$ & $0.116_{0.002}^{0.002}$ & $0.048_{0.002}^{0.002}$ & $0.035_{0.001}^{0.001}$ & $0.072_{0.002}^{0.002}$ & $0.057_{0.001}^{0.002}$\\
    ~  & $\sigma_c$ & ~ & $0.069_{0.004}^{0.004}$ & $0.041_{0.004}^{0.004}$ & $0.040_{0.004}^{0.004}$ & $0.105_{0.004}^{0.005}$ & $\llap{-}0.024_{0.004}^{0.006}$ & $\llap{-}0.034_{0.001}^{0.002}$ & $\llap{-}0.109_{0.002}^{0.002}$ & $\llap{-}0.024_{0.006}^{0.007}$\\
    \hline
    ~  & $\sigma_o$ & $0.128_{0.004}^{0.004}$ & ~ & $0.063_{0.002}^{0.003}$ & $0.153_{0.004}^{0.004}$ & $0.220_{0.006}^{0.005}$ & $0.099_{0.003}^{0.002}$ & $0.078_{0.002}^{0.002}$ & $0.071_{0.002}^{0.002}$ & $0.064_{0.002}^{0.002}$\\
    $R_{23.5}$ & $\sigma_b$ & $0.118_{0.003}^{0.003}$ & $\bullet$ & $0.063_{0.001}^{0.001}$ & $0.153_{0.003}^{0.001}$ & $0.210_{0.005}^{0.005}$ & $0.094_{0.002}^{0.002}$ & $0.073_{0.002}^{0.002}$ & $0.057_{0.002}^{0.002}$ & $0.061_{0.001}^{0.002}$\\
    ~  & $\sigma_c$ & $0.092_{0.005}^{0.005}$ & ~ & $\llap{-}0.069_{0.001}^{0.003}$ & $0.115_{0.005}^{0.006}$ & $0.201_{0.007}^{0.006}$ & $\llap{-}0.018_{0.014}^{0.030}$ & $\llap{-}0.033_{0.005}^{0.006}$ & $\llap{-}0.172_{0.001}^{0.001}$ & $\llap{-}0.082_{0.002}^{0.001}$\\
    \hline
    ~  & $\sigma_o$ & $0.233_{0.007}^{0.008}$ & $0.152_{0.004}^{0.007}$ & ~ & $0.234_{0.005}^{0.010}$ & $0.417_{0.015}^{0.012}$ & $0.115_{0.005}^{0.002}$ & $0.131_{0.004}^{0.004}$ & $0.196_{0.007}^{0.004}$ & $0.110_{0.003}^{0.003}$\\
    $L_{23.5}$ & $\sigma_b$ & $0.186_{0.006}^{0.006}$ & $0.152_{0.002}^{0.001}$ & $\bullet$ & $0.231_{0.004}^{0.003}$ & $0.392_{0.010}^{0.009}$ & $0.100_{0.003}^{0.003}$ & $0.107_{0.004}^{0.003}$ & $0.159_{0.005}^{0.005}$ & $0.088_{0.003}^{0.003}$\\
    ~  & $\sigma_c$ & $0.131_{0.012}^{0.013}$ & $\llap{-}0.165_{0.003}^{0.007}$ & ~ & $0.079_{0.015}^{0.025}$ & $0.374_{0.016}^{0.013}$ & $\llap{-}0.192_{0.003}^{0.001}$ & $\llap{-}0.123_{0.005}^{0.004}$ & $\llap{-}0.391_{0.004}^{0.002}$ & $\llap{-}0.203_{0.002}^{0.002}$\\
    \hline
    ~  & $\sigma_o$ & $0.306_{0.008}^{0.010}$ & $0.447_{0.011}^{0.012}$ & $0.294_{0.006}^{0.012}$ & ~ & $0.392_{0.012}^{0.012}$ & $0.231_{0.006}^{0.005}$ & $0.339_{0.011}^{0.010}$ & $0.478_{0.022}^{0.010}$ & $0.336_{0.014}^{0.007}$\\
    $\Sigma_{1}$ & $\sigma_b$ & $0.292_{0.007}^{0.007}$ & $0.423_{0.009}^{0.010}$ & $0.283_{0.007}^{0.006}$ & $\bullet$ & $0.363_{0.009}^{0.008}$ & $0.214_{0.005}^{0.005}$ & $0.324_{0.007}^{0.007}$ & $0.436_{0.010}^{0.010}$ & $0.316_{0.007}^{0.007}$\\
    ~  & $\sigma_c$ & $0.158_{0.016}^{0.017}$ & $0.336_{0.015}^{0.017}$ & $0.099_{0.019}^{0.032}$ & ~ & $0.296_{0.015}^{0.016}$ & $\llap{-}0.190_{0.007}^{0.007}$ & $0.230_{0.017}^{0.015}$ & $\llap{-}0.261_{0.037}^{0.019}$ & $0.136_{0.041}^{0.018}$\\
    \hline
    ~  & $\sigma_o$ & $0.185_{0.005}^{0.007}$ & $0.255_{0.008}^{0.006}$ & $0.224_{0.008}^{0.006}$ & $0.160_{0.005}^{0.005}$ & ~ & $0.164_{0.004}^{0.007}$ & $0.212_{0.006}^{0.008}$ & $0.261_{0.010}^{0.008}$ & $0.225_{0.007}^{0.006}$\\
    $g-z$ & $\sigma_b$ & $0.183_{0.003}^{0.003}$ & $0.240_{0.005}^{0.006}$ & $0.210_{0.005}^{0.005}$ & $0.150_{0.004}^{0.004}$ & $\bullet$ & $0.154_{0.003}^{0.004}$ & $0.209_{0.004}^{0.003}$ & $0.243_{0.006}^{0.006}$ & $0.211_{0.005}^{0.005}$\\
    ~  & $\sigma_c$ & $0.164_{0.006}^{0.008}$ & $0.233_{0.008}^{0.007}$ & $0.201_{0.009}^{0.007}$ & $0.121_{0.006}^{0.006}$ & ~ & $0.125_{0.005}^{0.009}$ & $0.194_{0.006}^{0.008}$ & $0.126_{0.021}^{0.017}$ & $0.195_{0.008}^{0.007}$\\
    \hline
    ~  & $\sigma_o$ & $0.248_{0.007}^{0.009}$ & $0.297_{0.010}^{0.007}$ & $0.144_{0.006}^{0.003}$ & $0.230_{0.006}^{0.005}$ & $0.399_{0.009}^{0.017}$ & ~ & $0.192_{0.007}^{0.006}$ & $0.306_{0.009}^{0.009}$ & $0.185_{0.007}^{0.006}$\\
    $M_{*}$ & $\sigma_b$ & $0.199_{0.007}^{0.006}$ & $0.275_{0.006}^{0.007}$ & $0.126_{0.003}^{0.004}$ & $0.220_{0.005}^{0.005}$ & $0.380_{0.009}^{0.009}$ & $\bullet$ & $0.170_{0.005}^{0.004}$ & $0.261_{0.008}^{0.007}$ & $0.163_{0.004}^{0.005}$\\
    ~  & $\sigma_c$ & $\llap{-}0.095_{0.016}^{0.026}$ & $\llap{-}0.053_{0.041}^{0.092}$ & $\llap{-}0.240_{0.003}^{0.002}$ & $\llap{-}0.190_{0.007}^{0.007}$ & $0.304_{0.012}^{0.021}$ & ~ & $\llap{-}0.163_{0.008}^{0.007}$ & $\llap{-}0.466_{0.006}^{0.006}$ & $\llap{-}0.248_{0.005}^{0.005}$\\
    \hline
    ~  & $\sigma_o$ & $0.125_{0.003}^{0.004}$ & $0.185_{0.005}^{0.006}$ & $0.131_{0.004}^{0.004}$ & $0.269_{0.009}^{0.008}$ & $0.394_{0.011}^{0.014}$ & $0.153_{0.006}^{0.005}$ & ~ & $0.162_{0.006}^{0.006}$ & $0.102_{0.004}^{0.002}$\\
    $M_{\rm dyn}$ & $\sigma_b$ & $0.113_{0.003}^{0.003}$ & $0.168_{0.005}^{0.004}$ & $0.105_{0.003}^{0.003}$ & $0.260_{0.006}^{0.005}$ & $0.386_{0.007}^{0.007}$ & $0.133_{0.004}^{0.004}$ & $\bullet$ & $0.139_{0.004}^{0.004}$ & $0.092_{0.002}^{0.002}$\\
    ~  & $\sigma_c$ & $\llap{-}0.109_{0.004}^{0.005}$ & $\llap{-}0.079_{0.012}^{0.014}$ & $\llap{-}0.123_{0.005}^{0.004}$ & $0.183_{0.013}^{0.012}$ & $0.361_{0.011}^{0.016}$ & $\llap{-}0.130_{0.007}^{0.006}$ & ~ & $\llap{-}0.392_{0.003}^{0.002}$ & $\llap{-}0.182_{0.003}^{0.002}$\\
    \hline
    ~  & $\sigma_o$ & $0.169_{0.007}^{0.004}$ & $0.114_{0.004}^{0.004}$ & $0.130_{0.005}^{0.003}$ & $0.261_{0.012}^{0.005}$ & $0.326_{0.012}^{0.011}$ & $0.163_{0.005}^{0.005}$ & $0.109_{0.004}^{0.004}$ & ~ & $0.106_{0.005}^{0.003}$\\
    j$_{*}$ & $\sigma_b$ & $0.156_{0.004}^{0.004}$ & $0.090_{0.003}^{0.003}$ & $0.105_{0.003}^{0.004}$ & $0.246_{0.006}^{0.005}$ & $0.305_{0.007}^{0.008}$ & $0.139_{0.004}^{0.004}$ & $0.093_{0.003}^{0.003}$ & $\bullet$ & $0.092_{0.003}^{0.002}$\\
    ~  & $\sigma_c$ & $\llap{-}0.232_{0.005}^{0.003}$ & $\llap{-}0.274_{0.002}^{0.002}$ & $\llap{-}0.260_{0.002}^{0.002}$ & $\llap{-}0.142_{0.020}^{0.011}$ & $0.158_{0.027}^{0.021}$ & $\llap{-}0.248_{0.003}^{0.003}$ & $\llap{-}0.262_{0.002}^{0.002}$ & ~ & $\llap{-}0.280_{0.002}^{0.001}$\\
    \hline
    ~  & $\sigma_o$ & $0.167_{0.006}^{0.005}$ & $0.124_{0.005}^{0.003}$ & $0.090_{0.002}^{0.003}$ & $0.218_{0.009}^{0.005}$ & $0.344_{0.010}^{0.009}$ & $0.120_{0.004}^{0.004}$ & $0.083_{0.003}^{0.002}$ & $0.130_{0.006}^{0.004}$ & ~\\
    j$_{\rm dyn}$ & $\sigma_b$ & $0.148_{0.004}^{0.004}$ & $0.115_{0.003}^{0.003}$ & $0.070_{0.002}^{0.002}$ & $0.207_{0.005}^{0.005}$ & $0.322_{0.008}^{0.007}$ & $0.105_{0.003}^{0.003}$ & $0.075_{0.002}^{0.002}$ & $0.110_{0.003}^{0.003}$ & $\bullet$\\
    ~  & $\sigma_c$ & $\llap{-}0.063_{0.015}^{0.017}$ & $\llap{-}0.159_{0.004}^{0.003}$ & $\llap{-}0.164_{0.002}^{0.002}$ & $0.088_{0.025}^{0.012}$ & $0.297_{0.011}^{0.010}$ & $\llap{-}0.161_{0.003}^{0.003}$ & $\llap{-}0.148_{0.002}^{0.002}$ & $\llap{-}0.341_{0.002}^{0.002}$ & ~\\
\enddata
\tablecomments{This table is formatted like \Fig{TriangleParameters} with results for each relation at the intersection of two variables. Each row gives every possible $x$-axis combination for a given $y$-axis. $\sigma_o$ is the observed forward scatter, $\sigma_b$ is the Bayesian intrinsic scatter (see \Sec{bayesianintrinsicscatter}), and $\sigma_c$ is the classical intrinsic scatter (see \Sec{classicalintrinsicscatter}). Uncertainty estimates for $\sigma_o$ and $\sigma_c$ are determined by bootstrap sampling; for $\sigma_b$, the uncertainty is taken from the posterior as the \wunits{68.3}{\%} credible interval. The superscript number is the positive $1\sigma$ value, and the subscript is the negative $1\sigma$ value. The diagonal cells are left empty.}
\end{deluxetable*}
\end{center}

The intrinsic scatters for all scaling relations under consideration are given in \Tab{scalingrelationscatters} in a format similar to \Tab{scalingrelationfits}.
All scatters are reported as the forward scatter for each parameter combination.
Observed scatters ($\sigma_o$) are computed as half the \wunits{15.9}{\%} to \wunits{84.1}{\%} quartile range of the residuals, which is similar to a standard deviation except more robust to outliers.
Bayesian scatters ($\sigma_b$) are computed following \Secs{bayesianintrinsicscatter}{numericalbayes} with uncertainty determined from the posterior.
Classical scatters ($\sigma_c$) are computed following \Sec{classicalintrinsicscatter} with uncertainty determined by the bootstrap method and taking the \wunits{15.9}{\%} to \wunits{84.1}{\%} quartiles to get asymmetric uncertainties.

Combining \Tabs{scalingrelationfits}{scalingrelationscatters} yields an orthogonal scatter via $\sigma_{\rm orthogonal} = \sigma_{\rm forward}/\sqrt{1+m^2}$.
The Bayesian intrinsic orthogonal scatter can be used to determine the tightest relation for each variable; however, some tight relations can be considered trivial if the variables depend on each other. 
The relations $V_{23.5} - M_{\rm dyn}$, $R_{23.5} - M_{\rm dyn}$, $(g-z) - M_{*}$, $(g-z) - \Sigma_{1}$, and $L_{23.5} - M_{*}$ are deemed trivial in that sense. 
The tightest relations for each parameter are then found to be: 
$V_{23.5} - M_{*}$;
$R_{23.5} - L_{23.5}, j_{*}, j_{\rm dyn}$; 
$L_{23.5} - V_{23.5}, R_{23.5}, j_{\rm dyn}$; 
$\Sigma_{1} - V_{23.5}$; 
$(g-z) - V_{23.5}$;
$M_{*} - V_{23.5}$; 
$M_{\rm dyn} - j_{\rm dyn}$;
$j_{*} - R_{23.5}$;
and $j_{\rm dyn} - V_{23.5}, R_{23.5}, L_{23.5}, M_{\rm dyn}$, 
with some parameters having multiple equally tight relations. 
It is perhaps not surprising that the TFR and stellar TFR are on the list of tightest relations. 
However, the other tight relations on this list also warrant a closer examination.

Another interesting aspect of \Tab{scalingrelationscatters} is the fraction of total scatter, which is intrinsic.  
The intrinsic scatter fraction is computed in quadrature as $f = \sigma_{\rm intrinsic}^2/\sigma_{\rm total}^2$ where our Bayesian results are used for the intrinsic scatter.
Most relations have intrinsic scatters ranging from 70 to 90 percent, with a few relations spanning greater extremes.
The $R_{23.5}-L_{23.5}$, $R_{23.5}-j_{\rm dyn}$, and $V_{23.5}- g-z$ relations, also found in our list of tightest scaling relations, have nearly 100 percent intrinsic scatter, making them ideally suited for comparisons with galaxy formation models. 

\begin{figure*}[ht]
  \centering
  \includegraphics[width=0.7\textwidth]{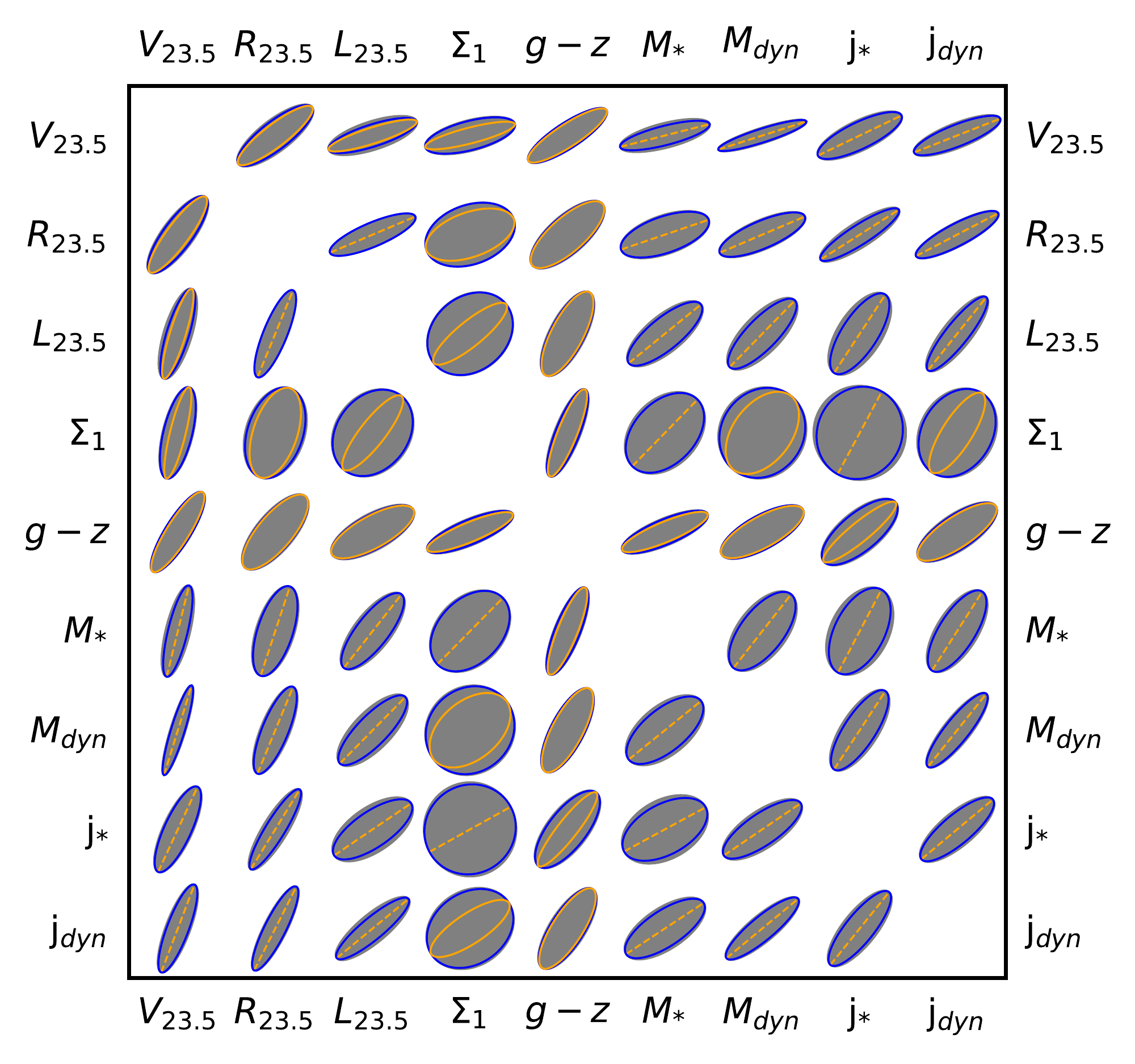}
  \caption{This figure is formatted like \Fig{TriangleParameters} to visualize the slope data from \Tab{scalingrelationfits} and the scatter data from \Tab{scalingrelationscatters}. 
  The gray, blue, and orange ellipses represent, respectively, the total observed scatter ($\sigma_o$), the Bayesian intrinsic scatter ($\sigma_b$), and the classical intrinsic scatter ($\sigma_c$). An orange dashed line through the center implies a negative classical intrinsic scatter. The thickness of each ellipse is proportional to the orthogonal scatter. Scatters associated with the $g-z$ colour are converted to dex to facilitate the comparison with other variables.}
  \label{fig:Figures/Ellipse_Compare_Bayesian.pdf}
\end{figure*}

To facilitate intuitive comparisons between scaling relations, \Fig{Figures/Ellipse_Compare_Bayesian.pdf} presents the slope data from \Tab{scalingrelationfits} and the scatter data from \Tab{scalingrelationscatters} graphically. 
The scatters are represented by ellipticity, which is set to the orthogonal scatter; thus, the forward and inverse versions of each relation are represented by an ellipse of the same thickness.
All axes are in dex except $g-z$ which is expressed in magnitude; the latter is converted into dex (dividing by 2.5) before plotting to allow for better visual comparison.
$\Sigma_1$ generally produces the broadest relations, though it does have a tight nontrivial relationship with $g-z$.

The orange dashed lines represent cases where the classical intrinsic scatter is negative, as is seen in several cases.
Most of these negative classical intrinsic scatter relations display a large covariance caused by one axis being strongly dependent on the other, the $V_{23.5} - M_{\rm dyn}$, $R_{23.5} - M_{\rm dyn}$, and $L_{23.5} - M_{*}$ relations being the clearest examples.
Other cases where the classical uncertainty is negative are due to a shared variable that dominates the uncertainty in each axis, such as the $R_{23.5} - L_{23.5}$, $R_{23.5} - M_{*}$, and $L_{23.5} - j_{\rm dyn}$ relations. 
These are all cases where the shared distance uncertainty dominates the covariance, rather than the dependence of one variable on the other.
While the possibility of a negative intrinsic scatter is a problematic aspect of a classical analysis, this can sometimes be rectified by identifying the shared uncertainty and simply omitting it from an uncertainty propagation.
More sinister are cases where the covariance is not strong enough to cause a negative intrinsic scatter and may remain unnoticed.
For example, all of the variables in this analysis share a covariance with inclination due to the internal extinction corrections (see \Sec{inclinationcorrections}; the rotational velocity correction also involves $1/\sin(i)$).

Several qualitative results become apparent upon inspection of \Tab{scalingrelationscatters} and \Fig{Figures/Ellipse_Compare_Bayesian.pdf}.
Foremost is that the Bayesian intrinsic scatters tend to be larger than the classical estimates.
This general result comes from the ability of the Bayesian method to account for covariant uncertainties, which \Fig{Perturbations} shows are very common.
All parameters, except $V_{23.5}$ and $g-z$, depend on distance and gain much of their observational uncertainty from distance errors; any relation with two distance-dependent parameters will thus have a large source of correlated error, which \App{testtruncation} shows as a critical point of failure for the classical method.
Without exception, inclination is a covariant source of uncertainty in all scaling relations studied in this work; though its effect is smaller than distance uncertainty, \App{testtruncation} indicates that even a small covariance can impact classical error propagations.
These covariant sources of error contribute less scatter to a relation than the naive classical method assumes. 
Therefore, the Bayesian method ultimately returns a higher intrinsic scatter value.
The effect is most pronounced for $M_{*}, M_{\rm dyn}, j_{*},$ and $j_{\rm dyn}$ where nearly all classical intrinsic scatter measurements for their scaling relations return negative values.
This is not too surprising as these are composite quantities that depend on many factors and so have many source of error to combine.

Another noticeable trend in \Tab{scalingrelationscatters} is that the \wunits{68.3}{\%} credible intervals for the Bayesian intrinsic scatter values are smaller than the \wunits{68.3}{\%} confidence intervals for the classical method.
This effect is most significant when the classical intrinsic scatter estimates approach (or cross) zero.
In the region around zero scatter, classical estimates are more unstable resulting in the larger uncertainty on the intrinsic scatter.
The behaviour of Bayesian intrinsic scatter estimates around zero is better defined, thanks in part to priors that prevent estimates from crossing that threshold.
Therefore, the Bayesian intrinsic scatter estimates will be biased high on average. 
However, they will also typically be closer to the true value than classical estimates (this is generally true of Bayesian methods, also see \App{testtruncation}).
An inadequate choice of model can also bias both Bayesian and classical intrinsic scatter estimates.
For instance, a poorly chosen inclination correction model will typically bias scatter measurements (and therefore intrinsic scatter estimates) higher than their true value.
These systematic effects are not reflected in the uncertainty ranges in \Tab{scalingrelationscatters} which only shows the random component.

Some specific cases in \Fig{Figures/Ellipse_Compare_Bayesian.pdf} stand out.
The $R_{23.5}$ and $L_{23.5}$ versus $V_{23.5}$ relations and most relations involving colour have similar intrinsic scatter predictions for the Bayesian and classical analyses.
These are the cases where most sources of uncertainty are not shared between each axis; 
it is no coincidence that $V_{23.5}$ and colour are in all such relations as they are distance independent (which is a major source of correlated uncertainty).
The relation for which the Bayesian and classical methods disagree is $g-z$ vs. $j_{*}$ which \Fig{Perturbations} shows has inclination as a large anti-correlated source of error.
Anticorrelated errors have a stronger impact on the Bayesian algorithm than the classical method.
The Bayesian algorithm cannot in fact assign a zero intrinsic scatter to a relation; 
one would instead have to perform a Bayes factor analysis to compare the zero and nonzero intrinsic scatter models.
This analysis would operationally be similar to our intrinsic scatter calculation, but a detailed description is beyond the scope of this paper.

\Fig{Figures/Ellipse_Compare_Bayesian.pdf} shows a comparison of many scaling relations by orthogonal scatter.
One can read across a row of the figure and see all cases with a single variable as the $y$-axis.
A discussion of the generalized applications of these relations, as standard candles for instance, is beyond the scope of this paper, though the format of \Fig{Figures/Ellipse_Compare_Bayesian.pdf} is conducive to exploring new relationships.

These broad trends indicate the necessity of using Bayesian intrinsic scatter measurements instead of classical techniques.
Differences between the two methods extend beyond the trivial second-order effect, but involve systematic and significant departures. 
We examine specific scaling relations more closely in \Sec{discussion}, and compare our results to the literature.

\section{Discussion}
\label{sec:discussion}

This section presents a comparison of some of our results with literature values. 
This exercise demands special attention since structural parameters are rarely measured in matching fashion from study to study. 
In \Tab{literaturecomparison} we present our literature comparisons for a few well-studied scaling relations as a subset of those relations presented in \Tabs{scalingrelationfits}{scalingrelationscatters}.
Whenever possible, we perform a unit conversion to units used in our analysis (for example $\log_{10}(L/[L_{\odot}])$ instead of magnitudes), detailing our transformations in the relevant subsection when these are nontrivial.
Bandpass transformations were not applied; instead, we report the published values and indicate the wavelengths in which the linear fit parameters were originally measured.  
Some parameters, such as luminosity, size, and colour, will be greatly affected by the choice of bandpass; others, such as stellar mass, central stellar surface density, and stellar angular momentum, are in principle bandpass independent.
However, systematic errors due to the choice of mass-to-light transformations may exist.
Comparing scaling relations based on bandpass-dependent quantities, such as the TFR, for different studies thus requires additional care.

Coupled with our analysis technique to evaluate intrinsic scatters, we can achieve higher precision in our estimates than previous (individual) analyses have allowed.
However, each relation deserves individual attention in order to fully realize its connection with galaxy formation and evolutionary models.
This section focuses on empirical (observational) results; detailed comparisons with simulations are beyond the scope of this paper.
The scaling relations examined briefly below set the stage for more detailed investigations elsewhere.

\startlongtable
\begin{deluxetable*}{c c c c c c c c c}
\tabletypesize{\scriptsize}
\tablewidth{\textwidth}
\tablecaption{Scaling Relation Literature Comparisons\label{tab:literaturecomparison}}
\tablehead{Source & $m$ & $\sigma_o$ & $\sigma_i$ & size & velocity & band & fit method & N \\
(1) & (2) & (3) & (4) & (5) & (6) & (7) & (8) & (9)} 
\tablecolumns{9}
\startdata\cutinhead{Tully-Fisher relation: ${\log_{10}(L/[L_{\odot}]) = m\log_{10}(V/[km\,s^{-1}]) + b}$}
This work & $3.191_{0.042}^{0.130}$ & $0.233_{0.007}^{0.008}$ & $0.186_{0.006}^{0.006}$ & \nodata & \Ha $V_{23.5}$ & $z$ & \citetalias{Akritas1996} bisector & 1152 \\
\citet{Tully1977} & $2.5\pm0.3$ & \nodata & \nodata & \nodata & \HI W & $B$ & visual & 18 \\
\citet{Pierce1988} & $3.14\pm0.12$ & 0.1 & $\leq 0.09$ & \nodata & \HI W & $I$ & OLS & 26 \\
\citet{Courteau1997} & $2.6\pm0.1$ & 0.2 & \nodata & \nodata & \Ha $V_{23}$ & $R$ & OLS & 304 \\
\citet{Verheijen2001} & $4.16\pm0.16$ & 0.12 & $0.05\pm0.05$ & \nodata & \HI $V_{flat}$ & $I$ & IOLS & 21 \\
\citet{Pizagno2005} & $2.603\pm0.133$ & \nodata & $0.131\pm0.015$ & \nodata & \Ha $V_{2.2}$ & $i$ & OLS+$\sigma$ & 81 \\
\citet{McGaugh2005} & $3.48\pm0.17$ & $0.24$ & \nodata & \nodata & \HI $V_{flat}$ & $B$ & OLS & 181 \\
\citet{Courteau2007} & $3.44\pm0.05$ & $0.197$ & $0.13$ & \nodata & \Ha $V_{2.2}$ & $I$ & ODR & 1303 \\
\citet{Pizagno2007} & $2.6\pm0.1$ & \nodata & $0.17\pm0.02$ & \nodata & \Ha $V_{80}$ & $z$ & OLS+$\sigma$ & 162 \\
\citet{Avila-Reese2008} & $3.83\pm0.18$ & 0.195 & $0.188$ & \nodata & \HI W & $K$ & ODR & 76 \\
\citet{Saintonge2011} & $3.63\pm0.01$ & $0.22$ & $0.14$ & \nodata & \HI W & $I$ & ODR & 3655 \\
\citet{Reyes2011} & $3.36\pm0.14$ & 0.22 & $0.16\pm0.03$ & \nodata & \Ha $V_{80}$ & $z$ & IOLS+$\sigma$ & 189 \\
\citet{Hall2012} & $3.66\pm0.09$ & 0.27 & 0.15 & \nodata & \HI W & $i$ & ODR & 668 \\
\citet{Bradford2016} & $3.40\pm0.05$ & $0.32\pm0.01$ & \nodata & \nodata & \HI W & $i$ & \citet{Kelly2007} & 930 \\
\citet{Ponomareva2017} & $3.25\pm0.24$ & $0.13\pm0.09$ & $0.14\pm0.03$ & \nodata & \HI $V_{flat}$ & $z$ & ODR & 32 \\
\citet{Ouellette2017} & $2.85\pm0.11$ & $0.20$ & $\leq 0.16$ & \nodata & \Ha $V_{23.5}$ & $i$ & OLS bisector & 46 \\
\cutinhead{Stellar Tully-Fisher relation: ${\log_{10}(M_*/[M_{\odot}]) = m\log_{10}(V/[km\,s^{-1}]) + b}$}
This work & $3.997_{0.039}^{0.144}$ & $0.248_{0.007}^{0.009}$ & $0.199_{0.007}^{0.006}$ & \nodata & \Ha $V_{23.5}$ & $z$ & \citetalias{Akritas1996} bisector & 1152 \\
\citet{Pizagno2005} & $3.048\pm0.121$ & \nodata & $0.158\pm0.021$ & \nodata & \Ha $V_{2.2}$ & $i$ & OLS+$\sigma$ & 81 \\
\citet{Avila-Reese2008} & $3.65\pm0.16$ & 0.21 & $0.16$ & \nodata & \HI W & $K$ & ODR & 76 \\
\citet{Dutton2010} & $3.56\pm0.04$ & \nodata & $0.18$ & \nodata & \Ha $V_{2.2}$ & $r$ & OLS & ${\sim}160$ \\
\citet{Reyes2011} & $3.60\pm0.13$ & 0.20 & $0.13\pm0.02$ & \nodata & \Ha $V_{80}$ & $i$ & OLS+$\sigma$ & 189 \\
\citet{Hall2012} & $3.79\pm0.14$ & 0.305 & 0.281 & \nodata & \HI W & $i$ & ODR & 668 \\
\citet{Bradford2016} & $4.16\pm0.06$ & $0.32\pm0.01$ & \nodata & \nodata & \HI W & SDSS & \citet{Kelly2007} & 930 \\
\citet{Ouellette2017} & $3.99\pm0.18$ & $0.32$ & \nodata & \nodata & \Ha $V_{23.5}$ & $i$ & OLS bisector & 46 \\
\citet{Lapi2018} & 3.42 & 0.08 & \nodata & \nodata & mixed $V_{3.2}$ & $I$ & OLS & 546 \\
\citet{Aquino-Ortiz2020} & $3.22\pm0.10$ & 0.20 & \nodata & \nodata & mixed $V_{max}$ & MaNGA & \citetalias{Akritas1996} & 200 \\
\cutinhead{Size-velocity relation: ${\log_{10}(R/[kpc]) = m\log_{10}(V/[km\,s^{-1}]) + b}$}
This work & $1.325_{0.034}^{0.074}$ & $0.128_{0.004}^{0.004}$ & $0.118_{0.003}^{0.003}$ & $R_{23.5}$ & \Ha $V_{23.5}$ & $z$ & \citetalias{Akritas1996} bisector & 1152 \\
\citet{Courteau2007} & $1.10\pm0.12$ & 0.17 & 0.15 & $R_{2.2}$ & \Ha $V_{2.2}$ & $I$ & ODR & 1303 \\
\citet{Avila-Reese2008} & $1.87\pm0.30$ & 0.290 & $0.285$ & $R_{d}$ & \HI W & $K$ & ODR & 76 \\
\citet{Saintonge2011} & $1.357\pm0.004$ & 0.11 & $0.084\pm0.001$ & $R_{23.5}$ & \HI W & $I$ & ODR & 3655 \\
\citet{Hall2012} & $1.518\pm0.065$ & 0.152 & 0.146 & $R_{23.5}$ & \HI W & $i$ & ODR & 668 \\
\citet{Ouellette2017} & $1.1\pm0.1$ & 0.137 & \nodata & $R_{23.5}$ & \Ha $V_{23.5}$ & $i$ & OLS bisector & 69 \\
\citet{Lapi2018} & 1.04 & 0.04 & \nodata & $R_{d}$ & mixed $V_{3.2}$ & $I$ & OLS & 546 \\
\cutinhead{Size-luminosity relation: ${\log_{10}(R/[kpc]) = m\log_{10}(L/[L_{\odot}]) + b}$}
This work & $0.417_{0.004}^{0.008}$ & $0.063_{0.002}^{0.003}$ & $0.063_{0.001}^{0.001}$ & $R_{23.5}$ & \nodata & $z$ & \citetalias{Akritas1996} bisector & 1152 \\
\citet{Courteau2007} & $0.32\pm0.01$ & 0.14 & 0.13 & $R_{2.2}$ & \nodata & $I$ & ODR & 1303 \\
\citet{Avila-Reese2008} & $0.285\pm0.033$ & 0.201 & $0.194$ & $R_{d}$ & \nodata & $K$ & ODR & 76 \\
\citet{Saintonge2011} & $0.413\pm0.003$ & 0.05 & $0.034\pm0.001$ & $R_{23.5}$ & \nodata & $I$ & ODR & 3655 \\
\citet{Hall2012} & $0.401\pm0.007$ & 0.076 & 0.060 & $R_{23.5}$ & \nodata & $i$ & ODR & 668 \\
\citet{Arora2021} & $0.36\pm0.01$ & $0.11\pm0.01$ & \nodata & $R_{23.5}$ & \nodata & $z$ & ODR & 2500 \\
\cutinhead{Size-stellar mass relation: ${\log_{10}(R/[kpc]) = m\log_{10}(M_*/[M_{\odot}]) + b}$}
This work & $0.334_{0.004}^{0.009}$ & $0.099_{0.003}^{0.002}$ & $0.094_{0.002}^{0.002}$ & $R_{23.5}$ & \nodata & $z$ & \citetalias{Akritas1996} bisector & 1152 \\
\citet{Shen2003} & $0.15/0.4$ & $0.20/0.15$ & \nodata & $R_{e}$ & \nodata & $z$ & OLS & 99,786 \\
\citet{Pizagno2005} & $0.242\pm0.030$ & \nodata & $0.142\pm0.011$ & $R_{d}$ & \nodata & $i$ & OLS+$\sigma$ & 81 \\
\citet{Fernandez2013} & $0.54/0.46/0.35$ & $0.12/0.12/0.12$ & \nodata & $R_{e}$ & \nodata & SDSS & OLS & $<452$ \\
\citet{Lange2015} & $0.21\pm0.02$ & \nodata & \nodata & $R_{e}$ & \nodata & $z$ & OLS & 6151 \\
\citet{Ouellette2017} & $0.34\pm0.02$ & 0.15 & \nodata & $R_{23.5}$ & \nodata & $i$ & OLS bisector & 69 \\
\citet{Lapi2018} & 0.23 & 0.05 & \nodata & $R_{e}$ & \nodata & $I$ & OLS & 546 \\
\citet{Wu2020} & $0.29_{0.07}^{0.06}$ & $0.2$ & \nodata & $R_{e}$ & \nodata & F814W & \nodata & $<1550$ \\
\citet{Trujillo2020Erratum} & $0.318\pm0.014$ & $0.087\pm0.005$ & $0.070\pm0.006$ & $R_{23.5}$ & \nodata & $i$ & \nodata & 464 \\
\citet{Arora2021} & $0.38\pm0.01$ & $0.11\pm0.01$ & \nodata & $R_{23.5}$ & \nodata & $z$ & ODR & 2433 \\
\cutinhead{Stellar-to-halo-mass relation: ${\log_{10}(M_*/[M_{\odot}]) = m\log_{10}(M_{\rm dyn}/[M_{\odot}]) + b}$}
This work & $1.256_{0.014}^{0.034}$ & $0.192_{0.007}^{0.006}$ & $0.170_{0.005}^{0.004}$ & $R_{23.5}$ & \Ha $V_{23.5}$ & $z$ & \citetalias{Akritas1996} bisector & 1152 \\
\citet{Reyes2011} & $1.28\pm0.06$ & 0.26 & $0.22\pm0.02$ & $R_{80}$ & \Ha $V_{80}$ & $i$ & OLS+$\sigma$ & 189 \\
\citet{Ouellette2017} & $1.27\pm0.07$ & 0.342 & 0.31 & $R_{23.5}$ & \Ha $V_{23.5}$ & $i$ & OLS bisector & 69 \\
\citet{Lapi2018} & 1.08 & 0.08 & \nodata & $R_{3.2}$ & mixed $V_{3.2}$ & $I$ & OLS & 546 \\
\cutinhead{${\Sigma_1}$-Stellar mass relation: ${\log_{10}(\Sigma_1/[M_{\odot}\,kpc^{-2}]) = m\log_{10}(M_*/[M_{\odot}]) + b}$}
This work & $1.004_{0.035}^{0.021}$ & $0.231_{0.006}^{0.005}$ & $0.214_{0.005}^{0.005}$ & \nodata & \nodata & $z$ & \citetalias{Akritas1996} bisector & 1152 \\
\citet{Barro2017} & $0.89\pm0.03$ & 0.25 & \nodata & \nodata & \nodata & CANDELS & \nodata & 1328 \\
\citet{Woo2019} & 0.86 & 0.24 & \nodata & \nodata & \nodata & $i$ & OLS & ${\sim}41000$ \\
\citet{Chen2020b} & 0.93 & \nodata & \nodata & \nodata & \nodata & MaNGA & OLS & 3654 \\
\citet{Arora2021} & $0.96\pm0.01$ & $0.24\pm0.01$ & \nodata & \nodata & \nodata & $grz$ & ODR & 2433 \\
\cutinhead{Specific angular momentum-mass relation: ${\log_{10}(j_{\rm dyn}/[kpc\,km\,s^{-1}]) = m\log_{10}(M_{\rm dyn}/[M_{\odot}]) + b}$}
This work & $0.814_{0.010}^{0.011}$ & $0.083_{0.003}^{0.002}$ & $0.075_{0.002}^{0.002}$ & $R_{23.5}$ & \Ha $V_{23.5}$ & \nodata & \citetalias{Akritas1996} bisector & 1152 \\
\citet{Takase1967} & $0.79$ & \nodata & \nodata & $R_{\infty}$ & $V_{\infty}$ & \nodata & \nodata & 18 \\
\citet{Zasov1989} & $0.79\pm0.08$ & \nodata & \nodata & $R_{25}$ & \HI W, $V_{max}$ & \nodata & \nodata & 34 \\
\cutinhead{Specific stellar angular momentum-stellar mass relation: ${\log_{10}(j_*/[kpc\,km\,s^{-1}]) = m\log_{10}(M_*/[M_{\odot}]) + b}$}
This work & $0.533_{0.006}^{0.011}$ & $0.163_{0.005}^{0.005}$ & $0.139_{0.004}^{0.004}$ & $R_{23.5}$ & \nodata & $z$ & \citetalias{Akritas1996} bisector & 1152 \\
\citet{Romanowsky2012} & $0.53\pm0.05$ & 0.22 & \nodata & $R_{\infty}$ & \nodata & $r$ & OLS & 64 \\
\citet{Cortese2016} & $0.80\pm0.09$ & 0.18 & \nodata & $R_{e}$ & \nodata & SAMI & \citetalias{Robotham2015} & 86 \\
\citet{Posti2018} & $0.55$ & \nodata & 0.19 & $R_{max}$ & \nodata & $3.6\mu$m & ODR & 92 \\
\citet{Lapi2018} & 0.50 & 0.05 & \nodata & $R_{3.2}$ & \nodata & $I$ & OLS & 546 \\
\citet{Sweet2018} & $0.56\pm0.06$ & \nodata & \nodata & $>3R_{e}$ & \nodata & mixed & \citetalias{Robotham2015} & 91 \\
\citet{Mancera2020} & $0.53\pm0.02$ & \nodata & $0.17\pm0.01$ & $R_{last}$ & \nodata & $3.6\mu$m & ODR & 132 \\
\enddata
\tablecomments{Column (1) gives the literature source of the scaling relation. Column (2) reports the slope and its uncertainty. Columns (3) and (4) give the forward observed and intrinsic scatter, respectively. In some cases for Columns (2)-(4), the values were transformed from their published value for consistency to ensure uniform units/meaning. Column (5) gives the choice of size metric, if relevant for that relation. Column (6) gives the velocity metric, if relevant, where ``\HI W'' is the \HI line width, and the velocity subscripts $V_{23.5},V_{2.2}, V_{80},V_{flat},V_{max}$, refer to the \wunits{23.5}{mag\,arcsec$^{-2}$} isophotal radius, 2.2 disk scale lengths, \wunits{80}{\%} of total light, a flat average, and the maximum radius available, respectively. Column (7) gives the bandpass for each measurement, when relevant. 
In some cases, the original survey from which the photometry was taken is indicated instead of a single bandpass. Column (8) gives an indication of the fitting method. This should be taken as a broad category; see the original article for more information. Where ODR, OLS, IOLS, OLS$+\sigma$ refer to orthogonal distance regression, ordinary least-squares regression, inverse ordinary least-squares regression, and ordinary least-squares with nonzero intrinsic scatter, respectively. Column (9) gives the number of galaxies in each fit. Dots [$\ldots$] mean that no data are available. If multiple relations were available from a given source, the most relevant version was selected for comparison ($z$ band, local universe, measurements at $R_{23.5}$, bisector fit, late-type galaxies, etc.). This table cannot properly capture the wealth of information found in the literature. 
Our quoted slope for \citet{Lapi2018} uses the derivative at the pivot point of their fit, ultimately using the value ``$y_1$'' from their Table 1; see the original paper for details.}
\end{deluxetable*}

\subsection{Tully-Fisher Relation: \texorpdfstring{${\log_{10}(L/[L_{\odot}]) = m\log_{10}(V/[km\,s^{-1}]) + b}$}{log(L) = mlog(V) + b}}\label{sec:luminosityvelocity}

The TFR between the rotational velocity, $V_{rot}$, and total luminosity\footnote{The TFR is often expressed in terms of absolute magnitude; in this case a factor of $-2.5$ transforms between the slope measurements.}, $L_\lambda$, of a spiral galaxy has received considerable attention since its inception \citep[][to cite a few]{Tully1977,Pierce1992,Strauss1995,Mo1998,Verheijen2001,Courteau2007,Ferrero2020}. 
The TFR itself is known to exhibit a wide range of slopes, intercepts, and scatters, depending on the choice of wavelength, sample, and analysis techniques. 
In their original paper, \citet{Tully1977} used photographic photometry and \HI line widths for a sample of 18 galaxies to find a photographic ($B$ band equivalent) slope of $2.5\pm0.3$.  
\citet{Courteau1997} used $R$ band photometry of 304 Sc galaxies and examined many different measures of velocity with two large samples, finding a variety of slopes from $2.14\pm0.14$ up to $2.86\pm0.08$ and forward scatter measurements ranging from 0.14 to \wunits{0.26}{dex} depending on the adopted parameters.
Later work by \citet{Courteau2007} revealed only weak correlations of the TFR residuals with colour, morphological type, surface brightness, size, and concentration~\citep[see also][]{Zwaan1995,Courteau1999}. 
In their Appendix, \citet{Courteau2007} provide theoretical derivations for the TFR slopes, showing that it can range from three to four depending on various assumptions.

\citet{Verheijen2001} performed a detailed analysis of the TFR parameters against sample, velocity metric, and passband choice, 
ultimately suggesting that the intrinsic scatter may be zero.
Their use of a classical analysis, which systematically underestimates the true value, is likely responsible for their low intrinsic scatter estimate.

The study of 162 spiral galaxies by \citet{Pizagno2007} focused on the scatter of the TFR given combinations of SDSS $g,r,i$, and $z$ photometric bands; three definitions of global velocity; and variations on extinction corrections, quality flags, and weighting of data points.
They again found that the choice of bandpass and analysis method had profound effects on the slope and scatter.

In an attempt to settle one element of this variability, the study by \citet{Hall2012} examined various reference radii for the measurement of luminosity and velocity, consistently finding that the isophotal radius, $R_{23.5}$, measured at the SB level of 23.5 $i$ band mag\,arcsec$^{-2}$, yields some of the tightest VRL relations (based on a forward scatter analysis).

An extensive examination of bandpass effects on the TFR by \citet{Ponomareva2017} found that 3.6$\mu m$ produces the tightest relation (by orthogonal scatter).
Having only 32 galaxies in their sample, random errors were a challenge for \citet{Ponomareva2017} and they found all wavelengths at, or longer than, the $i$ band had the same orthogonal scatter to within $1\sigma$ of each other.

The TFR has been extensively studied, and the works presented in this section are but a small representation of the total literature.
Still, we have established general consistency between our analysis and previous, mostly smaller, studies.  
Some of the above studies have reported measurements for the intrinsic TFR scatter, highlighting its importance for extragalactic studies.
\citet{Rhee2000} noted that the true intrinsic scatter is likely larger than classical estimates.
With our method, we can finally quantify the effect of covariant uncertainties on the intrinsic scatter of the TFR.

\subsection{Stellar Tully-Fisher Relation: \texorpdfstring{${\log_{10}(M_*/[M_{\odot}]) = m\log_{10}(V/[km\,s^{-1}]) + b}$}{log(M*) = mlog(V) + b}}\label{sec:stellarmassvelocity}

Two variants of the TFR, the stellar and baryonic TFR (STFR and BTFR, respectively), involve a correlation between the rotational velocity and the stellar or total baryonic masses of a galaxy~\citep{McGaugh2000, Verheijen2001, McGaugh2005, Hall2012, Bradford2016, Ouellette2017}.
Note that PROBES has stellar masses, but gas masses are largely unavailable (at the time of writing); therefore, the BTFR is beyond the scope of our research. 
For a comprehensive review of the BTFR, see \citet{Bradford2016}.
Here we will only examine the STFR; all STFR values are presented in \Tab{literaturecomparison}.

An operational advantage of the STFR and BTFR over the basic TFR is the elimination of bandpass dependencies. 
In principle, estimates of the STFR and BTFR should not depend on the methods used to obtain stellar masses and velocities. 
Therefore, it is surprising that the STFR should exhibit as much, if not more, internal variation in slope and scatter measures than the TFR.
Systematic differences in choices of mass-to-light transformations may be at play. 

Three studies \citep{Hall2012,Bradford2016,Ouellette2017} have consistent slope measures to ours; however, their scatter values are considerably larger.
Differences between the STFR and BTFR are mostly seen at low masses, and sample selection (e.g. yielding varying proportions of low-mass galaxies) may explain these. 

\subsection{Size-Velocity Relation: \texorpdfstring{${\log_{10}(R/[kpc]) = m\log_{10}(V/[km\,s^{-1}]) + b}$}{log(R) = mlog(V) + b}}\label{sec:sizevelocity}

The size-velocity relation (hereafter $RV$) is another tight correlation that can be used as a distance indicator or to constrain galaxy formation models. 
The slope of the $RV$ relation is expected to be near one for dark-matter-dominated systems~\citep{Mo1998,Courteau2007}. 
The PROBES sample includes all of the galaxies studied in \citet{Courteau2007}.  
Despite this significant overlap, we find a larger slope and smaller scatter.
These differences in $RV$ slope and scatter result mostly from using disk scale lengths~\citep{Courteau2007} instead of isophotal radii (this study) and the choice of inclination correction (see \Sec{inclinationcorrections}).

\citet{Saintonge2011} studied the $RV$ relation for a large collection of Sc galaxies from the SFI++ sample~\citep{Springob2007}.
Despite the close agreement in slopes, our scatter is considerably larger than theirs. 
The differences may stem largely from sample selection, with the PROBES sample covering a broader range of morphologies, as well as our velocity measures ($V_{23.5}$ versus \HI line width) and photometric band passes ($z$ versus $I$ band).

\citet{Hall2012} also explored combinations of velocity, size, and luminosity for 3041 spiral galaxies with \HI line widths and SDSS imaging.  
Their analysis considered different radial definitions, wavelength bands, and sample selection criteria, finding that structural parameters measured at $R_{23.5}$ (in their case for the $i$ band), yield the tightest scaling relations.
In \Tab{literaturecomparison}, we report the results from their highest-quality ``sample D,'' which has moderate inclinations and well-determined distances.
For their less restrictive ``sample B,'' they find an $RV$ slope of $1.334\pm0.046$, which matches our value more closely.  
Their use of integrated line widths instead of spatially resolved rotation curves should not significantly impact the scatter of the TFR \citep{Courteau1997}. 
However, resolved rotation curves inform us about the location of the velocity measurement whereas that spatial information is lost with line widths. 
Other potential sources of discrepancy include the choice of inclination corrections, and their distance calibration (see \citealt{Hall2012} Figure 19). 
Addressing the choice of distance calibration, \citet{Hall2012} found an $RV$ relation with slope 1.33 and scatter of \wunits{0.097}{dex}.
An intrinsic $RV$ scatter for their study can be inferred by taking their reported average uncertainties in size (\wunits{0.03}{dex}) and velocity (\wunits{0.02}{dex}) and performing a classical error analysis; this yields an intrinsic scatter of \wunits{0.09}{dex}, which is somewhat lower than our intrinsic scatter estimate.
Since this value agrees with our classical intrinsic scatter value, the discrepancy is likely due to expected differences between classical and Bayesian scatter analyses. 

The \citet{Lapi2018} analysis reported forward and inverse fits to their $RV$ relation.
Using the technique from \citet{Takashi1990}, we compute a bisector fit from the forward and inverse fits in \citet[][, Table 1]{Lapi2018}.
Their resulting slope is the shallowest among all reported in \Tab{literaturecomparison}. This discrepancy may stem from their use of a disk scale length as the size metric.
\citet{Courteau2007} also got a shallower slope and used disk scale lengths as their size metric.
The remarkably small scatter of \citet{Lapi2018} is likely connected to their use of stacked profiles.

\subsection{Size-Luminosity Relation: \texorpdfstring{${\log_{10}(R/[kpc]) = m\log_{10}(L/[L_{\odot}]) + b}$}{log(R) = mlog(L) + b}}\label{sec:sizeluminosity}

The size-luminosity relation (hereafter $RL$) has the interesting property that both axes are proportional to distance, and therefore distance errors will not significantly contribute to scatter.
Although useless as a distance indicator, the residuals of the $RL$ relation may offer useful comparisons with galaxy formation models without the nuisance of distance uncertainty \citep{Crain2015}.

There is a clear bimodality in $RL$ results, with \citet{Courteau2007} and \citet{Avila-Reese2008} getting shallow slopes and larger scatters than the other studies.
The use of disk scale lengths as the fiducial size metric in these two studies is the most likely source of discrepancy.
Sample size \citep[e.g.][]{Avila-Reese2008} can also be a factor. 
Studies that use $R_{23.5}$ as their size metric show relative agreement, though classical treatments \citep{Saintonge2011,Hall2012} underestimate the intrinsic scatter.

While \citet{Hall2012} did not compute an $RL$ intrinsic scatter directly, this can be inferred from their reported observational uncertainties. 
Their reported measurement errors for luminosity and size of \wunits{0.09}{dex} and \wunits{0.03}{dex}, respectively, imply an intrinsic scatter of \wunits{0.06}{dex} based on a classical analysis. 
Since our data and parameter extraction techniques are similar, the scatter difference is likely due to adopting a classical computation methodology. 
The classical analysis cannot account for covariances such as the shared effect of a photometric zero-point uncertainty (a positive fluctuation would increase both the luminosity and the size), the shared effect from distance uncertainties, and the shared effect of inclination uncertainties.
With so many shared variables, a classical analysis of the intrinsic scatter in the $RL$ relation is not possible without considering extra covariant terms. 

The $RL$ relation is an ideal testbed for the validation of intrinsic scatter analyses as the variables are strongly correlated, yet the relation itself is nontrivial. 
The scatter of this relation offers insight into the cosmic diversity of observed galaxies whilst minimizing the influence of observational uncertainties.
In fact, our Bayesian analysis finds that essentially all of the $RL$ scatter is intrinsic as uncertainty contributions are small.
It is also a highly reproducible relation, with our results and those of \citet{Saintonge2011} and \citet{Hall2012} being quite similar.

\subsection{Size-Stellar Mass Relation: \texorpdfstring{${\log_{10}(R/[kpc]) = m\log_{10}(M_*/[M_{\odot}]) + b}$}{log(R) = mlog(M*) + b}}\label{sec:sizestellarmass}

The size-stellar mass (hereafter $RM_*$) relation probes several aspects of galactic models including star formation history, angular momentum distribution, and coupling with the dark matter halo.
Given the dependence of $R$ and $M_*$ on distance, this scaling relation cannot be used as a distance estimator.
However, a distinct advantage of the $RM_*$ relation over the related $RL$ relation is the expected independence of $M_*$ on bandpass, making it simpler to compare across observational and numerical studies.
As stated previously, the choice of size metric acutely impacts the shape and scatter of the related scaling relations. 
Studies based on $R_{23.5}$ produce tighter $RM_*$ relations than other size metrics, and yield slopes consistent with ours ($\sim 0.33$), while those using $R_{d}$ or $R_{e}$ consistently get shallower slopes ($\sim 0.25$). 
Moreover, the intrinsic scatter from \citet{Pizagno2005} is almost twice our value.
While \citet{Trujillo2020Erratum} used $R_{23.5}$, their intrinsic scatter is slightly smaller than ours, as expected for a classical analysis.
As previously stated, \citet{Lapi2018} reported a tight $RM_*$ relation, despite using $R_{e}$, as a result of their stacking method. 
The explanation for the larger scatter found by \citet{Ouellette2017} could include small sample size and cluster environment (most other studies sample field environments).

\citet{Fernandez2013} considered morphological segregation in their analysis, including early-type galaxies.
We have only extracted their slopes for morphologies corresponding to Sb, Sbc, and Sc galaxies as these represent the best overlap with our samples.
\Tab{literaturecomparison} gives their $RM_*$ slopes and scatters separated by a ``/'' character.
\citet{Shen2003} also fitted two $RM_*$ relations, for low- and high-mass systems (with a transition mass of  $\sim 10^{10.5}~M_{\odot}$). 
Their use of a circular (Petrosian) half light radius for the relation however explains their extreme slopes and large scatter~\citep{Hall2012}. 

Overall, the small intrinsic scatter of the $RM_*$ relation offers an excellent benchmark for the comparison of observational and/or numerical studies.
However, results and their interpretation depend sensitively on the sample selection function and the adopted size metric.  

\subsection{Stellar-to-Total-Mass Relation: \texorpdfstring{${\log_{10}(M_*/[M_{\odot}]) = m\log_{10}(M_{\rm dyn}/[M_{\odot}]) + b}$}{log(M*) = mlog(Mdyn) + b}}\label{sec:stellarmasstotalmass}

The stellar-to-total-mass relation (hereafter STMR) plays an important role in understanding the interplay of baryons and dark matter; it is studied across a broad range of scales \citep[see, e.g.,][for a comparison of several studies]{Behroozi2013}.
Note that theoretical studies typically examine the Stellar-to-Halo Mass Relation, while observations are limited to STMRs within a finite galactocentric radius \citep{Ouellette2017}.
The slope of the STMR changes at a halo-mass scale of $\sim10^{12}\,M_{\odot}$~\citep{Moster2010}, which represents the star formation efficiency in massive halos.
The PROBES sample does not reach the turnover mass range, and we may fit a single power law to our data.

\citet{Reyes2011} and \citet{Ouellette2017} both report STMR slopes that are fully consistent with ours; however, their scatter and intrinsic scatter values are considerably larger. 
Both studies fit small samples, which increases the uncertainty on their results as a small number of outliers could greatly increase the observed scatter.
In the case of \citet{Reyes2011}, the increased scatter may also be due to their use of $R_{80}$ as the radius at which masses are measured, $R_{80}$ is a variant of the effective radius that is known to produce broader scaling relations.
In the case of \citet{Ouellette2017} our analysis techniques are well aligned and so the only other likely explanation is the environment.
The \citet{Ouellette2017} sample is taken from the Virgo cluster and our sample is a mix of environments with mostly field galaxies. 

\citet{Lapi2018} performed their analysis on stacked data and found an STMR slope and scatter that are considerably lower than ours.
Differences are likely explained by their use of stacked data and disk scale length as a size metric.

\subsection{\texorpdfstring{${\Sigma_1}$}{Sigma1}-Stellar Mass Relation: \texorpdfstring{${\log_{10}(\Sigma_1/[M_{\odot}\,kpc^{-2}]) = m\log_{10}(M_*/[M_{\odot}]) + b}$}{log(Sigma1) = mlog(M*) + b}}\label{sec:sigma1mstar}

The $\Sigma_{1}$-stellar mass relation is of great interest to the galaxy formation community.
$\Sigma_{1}$ has been empirically shown to be connected with black hole mass, and the transition to a quenched state~\citep{Chen2020}.
Unlike other measures such as stellar surface density within the effective radius ($\Sigma_{e}$), $\Sigma_{1}$ is insensitive to mergers~\citep{Szomoru2012,Barro2017}.
Stellar mass is of course sensitive to mergers, and so it is surprising that \citet{Barro2017} find the slope of the $\Sigma_{1}$-stellar mass relation to be constant as a function of redshift with a slope of $0.89\pm0.03$ (though the zero-point does evolve with time).
Our scatter measurements are in close agreement with those of \citet{Barro2017}, who reported a scatter of roughly \wunits{0.25}{dex} for their star forming galaxies sample, where our value is $0.231_{0.006}^{0.005}$ for a broad range of late-type galaxies.

\citet{Woo2019} fit an $\Sigma_{1}$ stellar mass and examine the position of galaxies in the relation and how it is connected to stellar age, specific star formation rate, and metallicity.
Specific evolutionary modes are indicated by the apparent paths that galaxies follow in this relation.
Fitting only star forming galaxies, they find a slope of 0.86, which is somewhat smaller than our value, though the difference may be attributed to the use of a least-squares fit on the relation.
Their scatter measurement of \wunits{0.24}{dex} is, however, in excellent agreement with our observed scatter.

\citet{Chen2020b} used a large sample of galaxies from the MaNGA~\citep{Bundy2015} survey to find stellar mass to be one of the strongest predictors of radial gradients in stellar population indicators.
We find excellent agreement with the \citet{Chen2020b} $\Sigma_{1}$-stellar mass slope of 0.93 for low-mass galaxies ($\log_{10}(M_{*}) < 10.95$); our reported slope is $1.004_{0.035}^{0.021}$.

\subsection{Specific Angular Momentum-Mass Relation: \texorpdfstring{${\log_{10}(j_{\rm dyn}/[kpc\,km\,s^{-1}]) = m\log_{10}(M_{\rm dyn}/[M_{\odot}]) + b}$}{log(jdyn) = mlog(Mdyn) + b}}\label{sec:jdynMdyn}

Dynamical angular momentum $J_{\rm dyn}$, like dynamical mass, is of special interest to galaxy formation studies since it is a conserved quantity (at least for isolated systems).
Angular momentum is also closely tied to dynamical mass through a simple theoretical prediction.
Assuming a galaxy is a solid rotating sphere, its moment of inertia will be $\frac{2}{5}MR^2$ and total mass is $M = \frac{4}{3}\pi\rho R^3$. 
Computing the angular momentum gives $J \propto M^{5/3}$; the specific angular momentum has $j \propto M^{2/3}$~\citep{Crampin1964}.
Galaxies are, of course, not solid rotating spheres, yet this assumption matches several observational studies, which we discuss below.

The two sources in \Tab{literaturecomparison} that report the $j_{\rm dyn} - M_{\rm dyn}$ relation have slopes that are consistent with ours; their scatter values are not available. 
The literature on $j_{\rm dyn} - M_{\rm dyn}$ relation is perhaps not as rich as other relations discussed above, perhaps because of the highly correlated nature of its two variables.
Our Bayesian analysis allows for a full accounting of correlations and their effect on intrinsic scatter estimates allowing us to confidently proceed with our analysis.
We find a tight relation with a robust slope considerably steeper the simple theoretical predictions above.
A possible interpretation is provided by considering the spin-parameter ($\lambda$) and energy ($E$) of a collapsing halo.  
This results in a similar power-law relation $j\propto M^{3/2}\lambda|E|^{-1/2}$, though with extra terms~\citep{Peebles1969}.
If Peebles's expression represents all relevant parameters, then we must have $\lambda|E|^{-1/2} \propto M^{-0.7}$ to roughly reproduce the observed relation.

\subsection{Specific Stellar Angular Momentum-Stellar Mass Relation: \texorpdfstring{${\log_{10}(j_*/[kpc\,km\,s^{-1}]) = m\log_{10}(M_*/[M_{\odot}]) + b}$}{log(j*) = mlog(M*) + b}}

The stellar version of the $j_{\rm dyn} - M_{\rm dyn}$ relation involves the complexities of baryonic physics.
Nevertheless, the uniformity of empirical results is remarkable (see \Tab{literaturecomparison}) with values for the slope and scatter consistent with ours. 

The relative consistency of slope and scatter estimates is quite remarkable given the range of size measures, photometric band passes, and sample sizes used in these various studies.
\citet{Lapi2018} consistently report very small scatters due to their stacking method.
Our intrinsic scatter is also smaller than those of \citet{Posti2018} and \citet{Mancera2020} 
possibly due to anticorrelated errors through the axial ratio (see \Fig{Perturbations}). 
Anticorrelated variables make the Bayesian intrinsic scatter estimates smaller than the classical ones.

Some studies of the $j_* - M_*$ relation include a third variable to account for the presence of a bulge~\citep{Obreschkow2014,Fall2018,Sweet2018}.
While our Bayesian intrinsic scatter algorithm is well suited for an arbitrary number of dimensions, a multidimensional analysis of scaling relations is beyond the scope of this paper.

\subsection{Future Directions}

Our analysis has revealed several avenues for future work.
The comparison of our results with other studies makes clear that sample selection is a major hindrance, as small samples do not produce reliable results; yet large samples may also exhibit a range of morphologies, masses, environments, etc that can significantly alter final results. 
Care is required to avoid sampling and selection bias. 

We have also (re)visited inclination corrections and their substantial effect on the slope and scatter of galaxy scaling relations.
This is discussed in some detail in \Sec{inclinationcorrections} and will undoubtedly require additional attention on many fronts (empirical, numerical, and theoretical).

Scaling relations also depend sensitively on the choice and definition of the parameters inherent to the relation itself. 
Most notably, the size metric at which quantities are measured can substantially affect the slope and scatter of the resulting scaling relations \citep{Courteau1997,Courteau2007,Hall2012,Bradford2016,Trujillo2020}.

Several scaling relations may have more complicated forms than a single power law, such as the curved STMR, and/or depend on a third parameter such as morphology, bulge/disk ratio, or stellar mass.
The Bayesian framework presented in \Sec{bayesianintrinsicscatter} is uniquely suited to address multidimensional analyses. 
Being entirely based on residuals, the Bayesian method is also well suited for the analysis of intrinsic scatter in nonlinear and multidimensional scaling relations.
Exciting avenues for future exploration that could benefit from our Bayesian intrinsic scatter analysis include the evolution of scaling relation scatter against a reference variable or with redshift. 
The same framework could also account for the changing scatter across a scaling relation, e.g., from the bright to the fainter end.
Being entirely forward modeling and residual based, our Bayesian intrinsic scatter analysis is powerful and flexible.

\section{Conclusions}
\label{sec:conclusions}

We have presented a Bayesian technique for computing intrinsic scatters of arbitrary scaling relations and demonstrated its robustness over the classical first-order method.
Because the Bayesian method relies  exclusively on forward parameter calculations, no derivatives or inverse functions are required to propagate measurement uncertainties.
We have explained the process of estimating Bayesian intrinsic scatters and compared them with ``classical'' estimates; the former is typically larger than the latter. 

We have also applied our method to a suite of observed galaxy scaling relations.
For these tests, we use the ``PROBES'' heterogeneous compilation of 1396 galaxies with spatially resolved \Ha rotation curves and homogeneous photometry extracted by us from the DESI-LIS.
Our scaling relations are constructed from all possible combinations of nine structural parameters for late-type galaxies.
The resulting scaling relations are homogeneously fit using a BCES bisector. 
Both Bayesian and classical intrinsic scatter values are then computed for each relation.
The agreement between our slope and scatter values and the literature is generally good (\Tab{literaturecomparison}). 

Our analysis has yielded the following data products and main results:
\begin{itemize}
    \item A robust set of homogeneously fit scaling relations from a large sample of late-type galaxies covering a broad range of structural properties (\Sec{results}). This includes tables of observed, classical, and Bayesian intrinsic scatters (\Tab{scalingrelationscatters}), and a detailed comparison with literature values (\Sec{discussion});
    \item A code to compute Bayesian intrinsic scatters (\App{bayesianintrinsicscattercode});
    \item An appreciation that Bayesian intrinsic scatters are typically larger than classical intrinsic scatters, as they account for parameter covariances and error propagation through robust fitting techniques (\Sec{intrinsicscatter}). Most relations have intrinsic scatters ranging from 60 to 90 percent of the total observed value, with a few relations spanning greater extremes. 
    \item An appreciation that Bayesian intrinsic scatters are more robust to biases in the data than classical methods (\App{testtruncation});
    \item An identification that the scatter of $R_{23.5}-L_{23.5}$, $V_{23.5} - (g-z)$, and $R_{23.5} - j_{\rm dyn}$ scaling relations is mostly intrinsic, largely because most observational errors slide along the scaling relations, making some of these relations ideally suited for comparisons with galaxy formation models;
    \item An identification of the tightest scaling relations by intrinsic orthogonal scatter (\Sec{intrinsicscatterresults}).
    For each structural parameter, these relations are:
    $V_{23.5} - M_{*}$;
    $R_{23.5} - L_{23.5}, j_{*}, j_{\rm dyn}$; 
    $L_{23.5} - V_{23.5}, R_{23.5}, j_{\rm dyn}$; 
    $\Sigma_{1} - V_{23.5}$; 
    $(g-z) - V_{23.5}$;
    $M_{*} - V_{23.5}$; 
    $M_{\rm dyn} - j_{\rm dyn}$;
    $j_{*} - R_{23.5}$;
    and $j_{\rm dyn} - V_{23.5}, R_{23.5}, L_{23.5}, M_{\rm dyn}$, 
    with some parameters having multiple equally tight relations. 
    \item A revised discussion on the nature and impact of projection corrections for galaxy structural parameters (\Sec{inclinationcorrections}) 
\end{itemize}

Genuine care is required in order to achieve meaningful analyses of galaxy scaling relations. 
As discussed in \Sec{discussion}, the parameters associated with any given galaxy scaling relation depend greatly on sample selection and analysis methods.
Accurate and representative structural parameters of galaxies, the simultaneous study of multiple scaling relations, and the derivation of robust measures of intrinsic scatter will facilitate comparisons with galaxy formation and evolutionary models.
The proper characterization of intrinsic scaling relation scatters ought to help bridge the gap between observations and simulations by removing one of the layers that stands between them.

\acknowledgments

We wish to acknowledge our Queen's University colleague Aaron Vincent for illuminating discussions about Bayesian statistics and data analysis, as well as the referee for insightful comments.  
Special thanks go to Arjun Dey and John Moustakas for discussions about the DESI Legacy Imaging Survey, and to the DESI team for the extensive database they provided.
This research has made use of the NASA/IPAC Extragalactic Database (NED),
which is operated by the Jet Propulsion Laboratory, California Institute of Technology, under contract with the National Aeronautics and Space Administration.
We are grateful to the Natural Sciences and Engineering Research Council of Canada, Ontario Government, and Queen's University for support through various scholarships and grants.

\software{astropy~\citep{astropy},
photutils~\citep{Bradley2020},
numpy~\citep{Harris2020}, 
scipy~\citep{scipy}}

\appendix

\section{Classical Uncertainty Functions}\label{app:classicaluncertainty}

Here, we report the classical uncertainty functions for all of the extracted parameters in \Sec{extractedparameters}.
These expressions are determined via \Equ{uncertaintyscatter} and include all sources of uncertainty from \Sec{datasets} for which we could assign a value and calculate derivatives.
\Tab{classicaluncertaintyequations} includes expressions for uncertainty for all parameters as used in our analysis.
   
\begin{center}
\begin{deluxetable*}{l l}
\tabletypesize{\normalsize}
\tablewidth{\textwidth}
\tablecaption{Classical Uncertainty Equations\label{tab:classicaluncertaintyequations}}
\tablehead{Variable & Classical uncertainty \\
(1) & (2)} 
\tablecolumns{2}
\startdata
        $\sigma_{\log_{10}(R_{23.5})}^2$ & $=\left(\frac{\sigma_{SB}}{R_{\rm obs}\Delta_{SB,R}\ln(10)}\right)^2 + \left(\frac{\sigma_{D}}{D\ln(10)}\right)^2 + \sigma_{C_{R_{23.5}}}^2$ \\
        $\sigma_{\log_{10}(L_{23.5,z})}^2$ & $=\left(\frac{\sigma_{m_z}}{2.5}\right)^2 + \left(2\frac{\sigma_{D}}{D\ln(10)}\right)^2 \left(\frac{\sigma_{R_{\rm obs}}\Delta_{m,R}}{2.5}\right)^2 + \sigma_{C_{L_{23.5}}}^2$\\ 
        $\sigma_{\log_{10}(V_{23.5})}^2$ & $=\frac{1}{\ln^2(10)}\left[\left(\frac{\sigma_{V_{\rm obs}}}{V_{\rm obs}}\right)^2 + \left(\frac{\sigma_{V_{\rm sys}}}{V_{\rm sys}}\right)^2 + \left(\frac{\sigma_i}{\tan(i)}\right)^2 + \left(\frac{\sigma_z}{1+z}\right)^2\right]$ \\
        $\sigma_{g-z}^2$ & $=\left(\sigma_{m_g}\right)^2 + \left(\sigma_{m_z}\right)^2 + \left(\sigma_{R_{\rm obs}}\Delta_{m_g,R}\right)^2 + \left(\sigma_{R_{\rm obs}}\Delta_{m_z,R}\right)^2 + \sigma_{C_{g-z}}^2$ \\
        $\sigma_{\log_{10}(M_{*})}^2$ & $=\frac{1}{\ln^2(10)}\left[ \left(\frac{\sigma_{L_{23.5}}}{L_{23.5}}\right)^2  + \left(\frac{\sigma_{g-z}\Delta_{\Upsilon,g-z}}{\Upsilon}\right) + \left(\frac{\sigma_{\Upsilon}}{\Upsilon}\right)^2\right]$ \\
        $\sigma_{\log_{10}(\Sigma_{1})}^2$ & $=\left(\frac{\sigma_{m_{1,z}}}{2.5}\right)^2 + \left(2\frac{\sigma_{D}}{D\ln(10)}\right)^2 + \sigma_{C_{\Sigma_{1}}}^2 + \left(\frac{\sigma_{\Upsilon_1}}{\Upsilon_1\ln(10)}\right)^2$ \\ 
        $\sigma_{\log_{10}(M_{\rm dyn})}^2$ & $=\left(\frac{2\sigma_{V_{23.5}}}{V_{23.5}\ln(10)}\right)^2 + \left(\frac{\sigma_{R_{23.5}}}{R_{23.5}\ln(10)}\right)^2$ \\
        $\sigma_{\log_{10}(j_{*})}^2$ & $=\sigma_{C_{j_{*}}}^2 + \sigma_{\log_{10}(M_{*})}^2 + \left(\frac{2\pi}{j_{*}\ln(10)}\right)^2\left[\left(\sigma_I\int_{0}^{R_{23.5}}\Upsilon(r)V(r)r^2dr\right)^2 + \left(\sigma_{\Upsilon}\int_{0}^{R_{23.5}}I(r)V(r)r^2dr\right)^2\right.$ \\
        ~ & $\left.+ \left(\sigma_{V_{23.5}}\int_{0}^{R_{23.5}}I(r)\Upsilon(r)r^2)dr\right)^2 + \left(\sigma_{R_{23.5}}I(R_{23.5})\Upsilon(R_{23.5})V(R_{23.5})R_{23.5}^2\right)^2\right]$ \\
        $\sigma_{\log_{10}(j_{\rm dyn})}^2$ & $=\sigma_{\log_{10}(M_{\rm dyn})}^2 + \left(\frac{\sigma_{V_{23.5}}}{j_{\rm dyn}}\int_{0}^{R_{23.5}}\frac{3V(r)^2r}{G\ln(10)}dr\right)^2 + \left(\frac{\sigma_{R_{23.5}}V(R_{23.5})^3R_{23.5}}{j_{\rm dyn}G\ln(10)}\right)^2$ \\
\enddata
\tablecomments{Expressions for classical uncertainty propagation as described in \Equ{uncertaintyscatter}. These are computed for all points in each scaling relation and used for the classical uncertainty propagation in \Sec{classicalintrinsicscatter}. The variables are described in the text and \Sec{extractedparameters}.}
\end{deluxetable*}
\end{center}

Some variable names appear in \Tab{classicaluncertaintyequations} for the first time and so we describe them here:
$\sigma_{D}$ is the uncertainty on the distance to a galaxy and appears in many of the relations.
$\sigma_{C_{R_{23.5}}}, \sigma_{C_{L_{23.5}}}, \sigma_{C_{g-z}}, \sigma_{C_{g-z}}, \sigma_{C_{\Sigma_{1}}},$ and $\sigma_{C_{j_{*}}}$ are the uncertainty on inclination correction factors for the corresponding variables. 
$\sigma_{SB}$ is the uncertainty in the surface brightness at the selected radius ($R_{\rm obs}$) and $\Delta_{SB,R}$ is the local slope of the SB profile at $R_{\rm obs}$. 
$\sigma_{m_z}$ is the uncertainty on the apparent $z$ band magnitude, and $\Delta_{m,R}$ is the slope of the curve of growth at the isophotal radius $R_{\rm obs}$. 
$\sigma_i$ is the inclination uncertainty (propagation through \Equ{inclination} not shown for clarity), and $\sigma_z$ is the redshift uncertainty.
$\sigma_{V_{\rm obs}}$ is the uncertainty on the individual velocity measurement at $R_{23.5}$, or the closest point to it. 
The uncertainty on the systematic velocity $\sigma_{V_{\rm sys}}$ was determined by fitting rotation curves with alternative methods (such as an arctan model) and finding typical agreement to within \wunits{3}{km s$^{-1}$}. 
This value is used as the systematic velocity uncertainty for all galaxies.
$\Delta_{m_g,R}$ and $\Delta_{m_z,R}$ are the slopes of the $g-$ and $z-$band growth curves, respectively.
$\Delta_{\Upsilon,g-z}$ is the slope of the mass-to-light ratio as a function of colour, and $\sigma_{\Upsilon}$ is the uncertainty on the mass-to-light ratio within $R_{\rm obs}$ set to \wunits{0.05}{dex} \citep[Sec 4.1 of][]{Roediger2015}.
Similarly, $\sigma_{\Upsilon_1}$ is the uncertainty on the mass-to-light ratio within \wunits{1}{kpc}.
$\sigma_{m_{1,z}}$ is the uncertainty on the magnitude within \wunits{1}{kpc}.
$\sigma_{I}$ is the uncertainty on the intensity at $R_{23.5}$.

\section{Testing Data Truncations}
\label{app:testtruncation}

\begin{figure}[ht]
  \centering
  \includegraphics[width=\columnwidth]{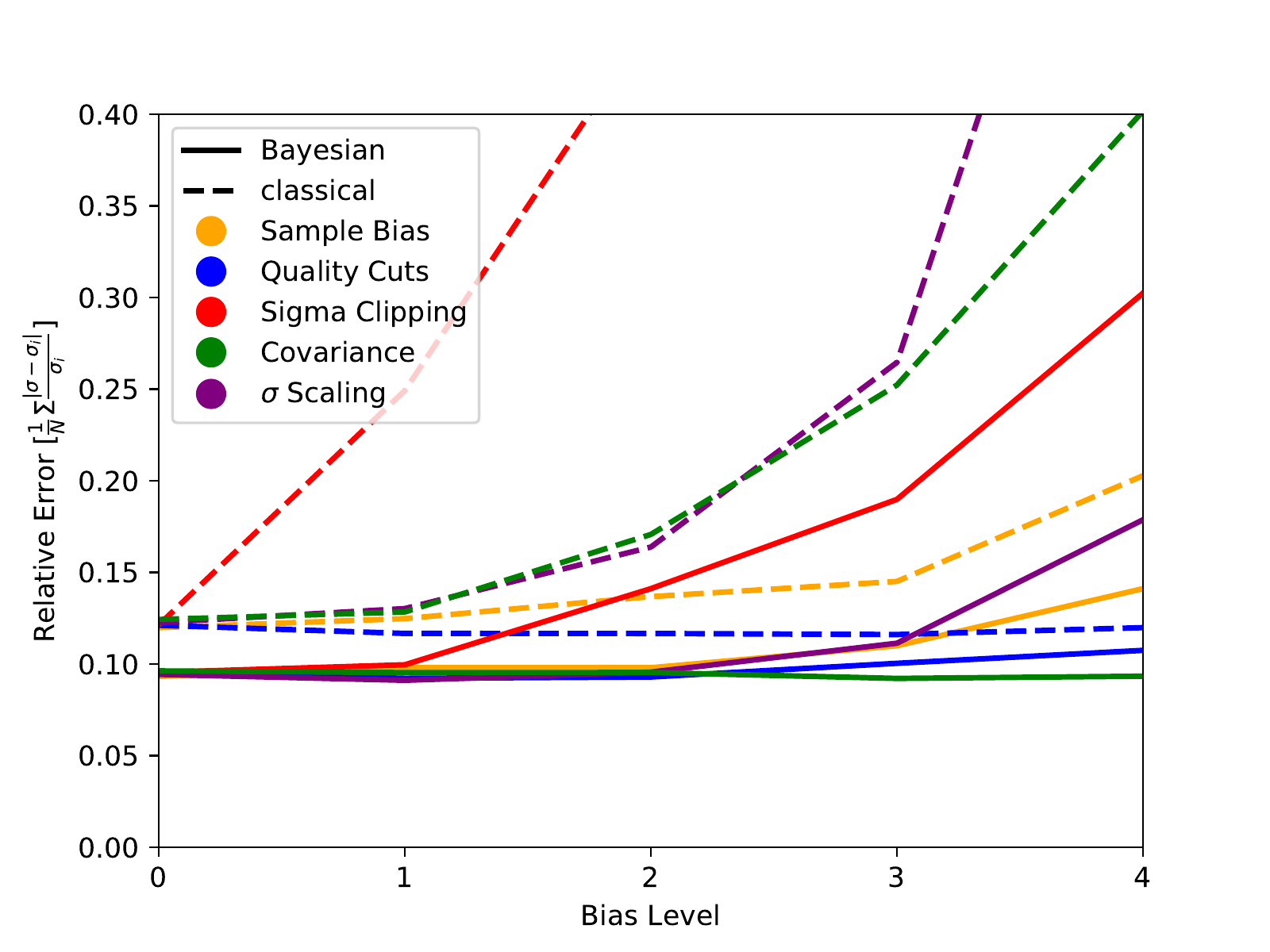}
  \caption{Tests on intrinsic scatter estimates, showing the relative error between each method and the true mock intrinsic scatter value averaged over 1000 trials.
    Shown are the relative errors for four levels of data manipulation as given in~\Tab{biases}; zero indicates no modification to the data. The Bayesian method consistently performs better than the classical method and is more robust to data manipulation.}
  \label{fig:biases}
\end{figure}

Various alterations of the toy model can be made to test the effects of realistic biases in the data.
Truncating the data on an axis can represent sampling biases. We test the effect of truncating extreme values on the $x$-axis by setting a lower bound to the axis that removes a given percentage of the data points.
Truncating based on uncertainty can represent data quality cuts, and we test the effect of setting an uncertainty threshold that removes some percentage of the data.
Truncating based on the residual after fitting a relation can represent sigma clipping, and we test the effect of removing data points with large residuals as well.
Shown in \Fig{biases} are tests for the stability of each method against five types of bias in the data; each test is performed using 1000 trials of the toy model.
We compare the relative error between the predicted intrinsic scatter value and the true value as a function of the degree of data bias.

\begin{center}
\begin{deluxetable}{l c c c c}
\tabletypesize{\normalsize}
\tablewidth{\columnwidth}
\tablecaption{Toy Model Modifications\label{tab:biases}}
\tablehead{Bias & 1 & 2 & 3 & 4 \\
(1) & (2) & (3) & (4) & (5)} 
\tablecolumns{5}
\startdata
    quality cuts & 0.05 & 0.10 & 0.25 & 0.50\\  
    sampling bias & 0.05 & 0.10 & 0.25 & 0.50\\ 
    sigma clipping & 0.01 & 0.02 & 0.03 & 0.05\\
    scaled & 0.02 & 0.05 & 0.10 & 0.25\\       
    covariance & 0.01 & 0.02 & 0.05 & 0.10\\   
\enddata
\tablecomments{Table of data modifications from \Fig{biases} to test the Bayesian and classical intrinsic scatter methods. Column (1) indicates the type of bias applied to the data. Columns (2) - (5) indicate the degree of bias. The numbers in the table have a different meaning for each bias type. Quality cuts, sampling bias, and sigma clipping values refer to the fraction of removed data points. Scaled refers to the factor by which the mock observation uncertainties are scaled relative to their true value. Covariance refers to the degree of extra covariant scatter introduced to the mock observations (note that the toy model intrinsic scatter is approximately 0.13 in the same arbitrary units).}
\end{deluxetable}
\end{center}

\Fig{biases} makes clear that scatter analyses are adversely impacted by sigma-clipping methods, since their accuracy and uncertainty bounds are invalidated.
This effect becomes significant for the both classical and Bayesian algorithms once as little as \wunits{1-2}{\%} of the data are removed. 
Indeed, sigma clipping eliminates some of the most informative data points from the sample; points close to the relation are consistent with any intrinsic scatter, while those far from the relation that have small uncertainty can only be explained with a nonzero intrinsic scatter.
\Fig{biases} also shows that truncations based on an axis or on uncertainty (as are used in \Sec{dataqualitycuts}) have essentially no effect on either method.
We tested truncating up to \wunits{50}{\%} of the data, and the relative error only increased slightly as would be expected with less data to analyze.
Finally, we see that the Bayesian method performs better than the classical method in the ideal toy model (zero bias) by about \wunits{25}{\%}.  
The difference is even more pronounced if there is some bias to the data such as correlated errors.

More sinister is the possibility that the observational uncertainties are themselves incorrect.
Perhaps a given uncertainty is overestimated as the observer wished to present conservative values, or the uncertainty could be underestimated due to an unknown factor influencing the data.
Whatever the cause, biased uncertainties can potentially present a significant barrier to measuring accurate intrinsic scatter values.
We test the effect of modifying the uncertainty quality.
We also test the impact of correlated uncertainties and verify the statement in \Sec{intrinsicscatter} that the Bayesian algorithm properly handles correlated uncertainties. 

To test for incorrect observational uncertainties, we randomly selected \wunits{50}{\%} of the data points and scaled their uncertainty by some factor, thus considering both inconsistent data (only half the data is modified) and incorrect uncertainty values.
This scaling is done after each data point is perturbed by its uncertainty, so the algorithms will only have access to the scaled (incorrect) values.

To test for correlated uncertainties, an extra uncertainty $\sigma_{z}$ is added to simultaneously scatter the data in both axes.
The classical method considers each axis separately and so the $\sigma_{z}$ error will be added in quadrature to both axes, while the Bayesian technique includes the new uncertainty in its sampling algorithm as a covariant term.

\Fig{biases} also shows that both algorithms can handle a small-scale manipulation to the uncertainty values; however the Bayesian algorithm estimate and its uncertainty are consistently more robust.
At \wunits{10}{\%} scaling, the classical algorithm produces an essentially meaningless uncertainty range.
The test for covariance in \Fig{biases} demonstrates, as expected, that the classical method is negatively impacted by this type of uncertainty while the Bayesian method handles it smoothly.

\section{Bayesian Intrinsic Scatter Code}
\label{app:bayesianintrinsicscattercode}

Here we present the Bayesian intrinsic scatter code.
The Python code below relies on the public numpy~\citep{Harris2020} and scipy~\citep{scipy} packages.
The code defines the function \texttt{BayesianIntrinsicScatter} which has several arguments described below.

The argument \texttt{phi\_list} is a list object where each element contains all of the measurement information for a galaxy (rotation curve, surface brightness profile, distance, etc.) and the associated uncertainties.
The galaxy object (elements of the \texttt{phi\_list} list) may be formatted in any way that is easiest for the user as the code does not directly interact with them, instead only passing them to other functions.
\texttt{sigma\_max} is the maximum possible intrinsic scatter value for the prior; typically the total scatter of the scaling relation is used.
The two \texttt{X} and \texttt{Y} arguments are functions that take a \texttt{phi\_list} element and evaluate the $x$/$y$ axes of the scaling relation in question.
These functions can return \texttt{None} to indicate that a point has exceeded a data quality cut and should be ignored.
Other than returning \texttt{None}, no other assumptions are made about the output of \texttt{X} and \texttt{Y} as they are only passed to user-defined functions.
\texttt{sample\_phi\_params} is a function that takes no arguments and returns any parameters that are needed for the \texttt{sample\_phi} function.
\texttt{sample\_phi} takes parameters from \texttt{sample\_phi\_params} and an element from \texttt{phi\_list} and returns the element resampled about its uncertainty values.
For example, the distance measurement would be resampled by a normal distribution about the original measurement with the uncertainty as the standard deviation.
\texttt{relation\_f} is the function for the scaling relation; it takes as arguments a tuple of parameters, an $x$-axis value, and a $y$-axis value, then returns the residual.
\texttt{relation\_f\_fit} takes a list of $x$-axis values and a list of $y$-axis values and returns the tuple of parameters needed for \texttt{relation\_f}.
\texttt{N\_samples} is the number of times to sample the \texttt{sample\_phi} function for each galaxy and should be at least 500 for the posterior to converge.
\texttt{N\_sigma} is the number of points to evaluate the intrinsic scatter pdf.
\texttt{sigma\_min} is the minimum value at which to evaluate the intrinsic scatter pdf.
\texttt{nprocs} is the number of processors used for the calculations; this can accelerate the Bayesian intrinsic scatter measurement.
If the speed up is not needed, one can replace the instances of \texttt{pool.map} with \texttt{map} for the same functionality.
\texttt{min\_pass} is the minimum number of evaluations out of \texttt{N\_samples} that do not return \texttt{None} required for a galaxy to be included in the intrinsic scatter calculation.

While the Bayesian intrinsic scatter calculation is somewhat more complicated than a classical intrinsic scatter analysis, there are some distinct advantages.
From a coding perspective, there are only a few functions that need to be defined.
The \texttt{X} and \texttt{Y} functions are effectively already needed to construct the scaling relation in the first place.
The \texttt{sample\_phi} function is relatively simple to construct as it typically only involves sampling normal distributions.
Again the \texttt{relation\_f} and \texttt{relation\_f\_fit} functions must already be constructed in order to fit the scaling relation.
Contrast this with the classical error analysis where derivatives must be computed for each axis as a function of each input value that has an uncertainty.
Once implemented for the first time, the Bayesian intrinsic scatter calculation is actually easier than the classical analysis, though it does take longer to compute.

{\scriptsize
\hrule\vspace{1pt}
\hrule
\begin{verbatim}
import numpy as np
from scipy.integrate import trapz
from scipy.stats import norm
from functools import partial
from multiprocessing import Pool

def _pdf(r,s):
    return np.sum(norm.pdf(r, loc = 0, scale = s))

def BayesianIntrinsicScatter(phi_list, sigma_max, X, Y,
                             sample_phi_params, sample_phi,
                             relation_f, relation_f_fit,
                             N_samples = 500, N_sigma = 100,
                             sigma_min = 0.01, nprocs = 4,
                             min_pass = 100):
    pool = Pool(nprocs)
    # array of values at which to evaluate the intrinsic scatter pdf
    S = np.linspace(sigma_min, sigma_max, N_sigma)

    residuals = [[] for n in range(len(phi_list))]
    for r in range(N_samples):
        params = sample_phi_params()
        # Resample galaxy list
        sample = pool.map(partial(sample_phi, params), phi_list)
        # Evaluate scaling relation axes
        XY = zip(pool.map(X,sample), pool.map(Y,sample),
                 range(len(phi_list)))
        XY = list(filter(lambda xy: not None in xy, XY))
        # Fit scaling relation
        fit = relation_f_fit(list(xy[0] for xy in XY),
                             list(xy[1] for xy in XY))
        for xy in XY:
            # Store scaling relation residual
            residuals[xy[2]].append(relation_f(fit, xy[0], xy[1]))
            
    posteriors = []
    for r in filter(lambda r: len(r) > min_pass, residuals):
        # evaluate the intrinsic scatter pdf for this galaxy
        pdf = np.array(pool.map(partial(_pdf, r), S))
        
        # normalize pdf to integral 1
        posteriors.append(np.log10(pdf/trapz(pdf, S)))
        
    # Take product of single galaxy posteriors and normalize
    P_sigmai = np.sum(posteriors, axis = 0)
    P_sigmai -= np.max(P_sigmai)
    P_sigmai = (10**P_sigmai)/trapz(10**P_sigmai, S)
    
    return S, P_sigmai
\end{verbatim}\hrule}

\bibliography{paper}

\end{document}